\documentclass[twocolumn,preprintnumbers,nofootinbib,prd,superscriptaddress,aps]{revtex4-1}

\usepackage[utf8]{inputenc}
\usepackage{graphicx,amssymb,amsmath,amsthm,amsfonts,epstopdf,epsfig,epsf,times}
\usepackage[linktocpage]{hyperref}
\usepackage[usenames]{color}
\usepackage{epstopdf}
\usepackage{physics}
\graphicspath{{plots/}} %setting the picture directory

\usepackage{bm}
\usepackage{dcolumn}
\usepackage{latexsym}
\usepackage{rotating}
\usepackage{hyperref}
\usepackage{color}
\usepackage{longtable}
\usepackage{enumerate}
\usepackage{tensor}
\usepackage{stmaryrd}
\usepackage[normalem]{ulem}
\usepackage{mathtools}
\usepackage{url}
\definecolor{coolblack}{rgb}{0.0, 0.18, 0.39}
\definecolor{darkred}{rgb}{0.5,0,0}
\definecolor{darkgreen}{rgb}{0,0.5,0}
\definecolor{darkblue}{rgb}{0,0,0.5}
\definecolor{lapislazuli}{rgb}{0.15, 0.38, 0.61}
\definecolor{venetianred}{rgb}{0.78, 0.03, 0.08}
\definecolor{bleudefrance}{rgb}{0.19, 0.55, 0.91}
\definecolor{dogwoodrose}{rgb}{0.84, 0.09, 0.41}
\definecolor{dogwoodrose}{rgb}{0.84, 0.09, 0.41}
\definecolor{darkorgane}{rgb}{1,0.549,0}
\hypersetup{colorlinks=true, citecolor=darkblue, linkcolor=darkblue,
    urlcolor = darkblue}

\DeclareMathOperator{\arcsinh}{arcsinh}
\definecolor{olive}{rgb}{0.5, 0.5, 0.0}

\newcommand{\ben}{\begin{enumerate}}
    \newcommand{\een}{\end{enumerate}}

\newcommand{\leqsim}{\,\mbox{{\scriptsize $\stackrel{<}{\sim}$}}\,}
\newcommand{\geqsim}{\,\mbox{{\scriptsize $\stackrel{>}{\sim}$}}\,}

\def\be{\begin{equation}}
    \def\ee{\end{equation}}
\newcommand{\beq}{\begin{eqnarray}}
    \newcommand{\eeq}{\end{eqnarray}}
\newcommand{\ba}{\begin{align}}
    \newcommand{\ea}{\end{align}}

\definecolor{darkorange}{rgb}{1,0.549,0}

\definecolor{tangerineyellow}{rgb}{1.0, 0.8, 0.0}
\definecolor{springgreen}{rgb}{0.0, 1.0, 0.5}
\definecolor{bostonuniversityred}{rgb}{0.8, 0.0, 0.0}

\def\be{\begin{equation}}
    \def\ee{\end{equation}}
    
\newcommand{\bea}{\begin{eqnarray}}
    \newcommand{\eea}{\end{eqnarray}}

\begin{document}\title {\large Two-body problem in theories with kinetic screening}
    
    \author{Mateja Bo\v{s}kovi\'{c}}
    \author{Enrico Barausse}
    \affiliation{SISSA, Via Bonomea 265, 34136 Trieste, Italy and INFN Sezione di Trieste}
    \affiliation{IFPU - Institute for Fundamental Physics of the Universe, Via Beirut 2, 34014 Trieste, Italy}

    \begin{abstract}
   	 New light scalar degrees of freedom may alleviate  the dark matter and dark energy problems, but if coupled to matter, they generally mediate a fifth force. In order for this fifth force to be consistent with  existing constraints, it must be suppressed close to matter sources, e.g. through a non-linear screening mechanism. In this work, we investigate the non-relativistic two-body problem in shift-symmetric scalar-tensor theories that exhibit kinetic screening ($k$-mouflage), both numerically and analytically.      
     We develop an approximate scheme, based on a Hodge-Helmholtz decomposition of the  Noether current associated to the shift symmetry, allowing for a qualitative insight into the dynamics and yielding results in good agreement with the numerical ones
     in most of the parameter space. We apply the formalism to polynomial $k$-essence and to Dirac-Born-Infeld (DBI) type theories, as well as to theories that develop ``anti-screening''. In the deep nonlinear regime, we find that the fifth force  is screened slightly more efficiently 
     in equal-mass systems than in extreme mass-ratio ones. However, we find that systems with comparable masses  also exhibit regions  where the screening is ineffective. These descreened spheroidal regions (bubbles) could in principle be probed in the solar system with sufficiently precise space accelerometers.
    \end{abstract}

    \date{\today}
    
    \maketitle
    
%    \tableofcontents

    %%%%%%%%%%%%%%%%%%%%%%%%%%%%%%%%%%%%%%%%%%%%%%
    \section{Introduction}\label{sec:intro}
    %%%%%%%%%%%%%%%%%%%%%%%%%%%%%%%%%%%%%%%%%%%%%%

    Probes from laboratory to cosmological scales are consistent with gravity being described by General Relativity (GR)~\cite{Will:2014kxa,Planck:2015bue,LIGOScientific:2021sio}. On the theoretical side, Poincar{\'e} invariance and unitarity at low energies [i.e. in the infra-red (IR)] imply that  the long-range attractive universal interaction that couples to both massive and massless particles must originate from  a spin-2 massless particle (the graviton), if one requires a single force carrier~\cite{Schwartz:2014sze}. In the classical limit, such a theory can be non-linearly completed to GR~\cite{Deser:1969wk,Deser:2009fq}. GR is a predictive effective field theory (EFT) up to the Planck scale, where it needs to be ultra-violet (UV) completed into a full theory of quantum gravity~\cite{Donoghue:2022eay}.
    
    On the IR side, there is also clear evidence for the presence of new forms of matter - dark energy (DE)~\cite{SupernovaSearchTeam:1998fmf,Planck:2018vyg, Brax:2017idh} and dark matter (DM)~\cite{Peebles:1982ff,Planck:2018vyg, Ivanov:2019pdj,Bertone:2016nfn}. As the presence of both of these components is inferred only from  gravitational experiments, it is reasonable to ask whether the gravitational interaction stems from other degrees of freedom, in addition to tensor gravitons, in order to explain these phenomena
    (partially or in full).

    For instance, cosmological observations are consistent with the presence of an effective cold DM component on large scales~\cite{Peebles:1982ff,Planck:2018vyg, Ivanov:2019pdj,Bertone:2016nfn}. There are a plethora of models providing a microphysical explanation for this  component, although none has yet been confirmed experimentally~\cite{Bertone:2016nfn,Bertone:2018krk}. An alternative approach, the phenomenological proposal of modifying Newtonian gravity, known as Modified Newtonian Dynamics (MOND)~\cite{Bekenstein:1984tv,Famaey:2011kh}, had some success on galactic scales~\cite{McGaugh:2016leg}. Possible relativistic generalizations of  MOND~\cite{Bekenstein:2004ne,Zlosnik:2006zu,Blanchet:2011wv, Bonetti:2015oda}  seem to be disfavored by LIGO/Virgo observations~\cite{Chesler:2017khz, Ezquiaga:2017ekz} and are yet unsuccessful  at explaining the structure of the Universe on large scales~\cite{Dodelson:2011qv}. However, some proposals (e.g. superfluid DM~\cite{Berezhiani:2015bqa}) have also attempted to combine cold DM's success on extra-galactic scales with  MOND's advantages on galactic scales. If these models are viable, a MOND-like phenomenology would  arise from the fifth force operating on galactic scales.
    
    Beyond-GR scalar-tensor theories may also provide an effective DE phenomenology on large cosmological scales. Examples
    include Horndeski theory~\cite{Horndeski:1974wa} and its generalizations~\cite{Crisostomi:2016czh, Creminelli:2017sry, Santoni:2018rrx, Brax:2017idh}, which can be organized in an ``EFT of DE''~\cite{Gubitosi:2012hu}. Some of these models also provide hints about how the cosmological constant problem could be addressed without resorting to an anthropic explanation~\cite{Brax:2017idh}.  The problem with such an approach is that one needs to explain the absence of fifth forces on local (e.g. solar system) scales.
    
    For example, the simplest way of extending GR is to introduce a massless scalar conformally coupled to matter, which leads to Fierz-Jordan-Brans-Dicke (FJBD) theory~\cite{Fierz:1956zz,Jordan:1959eg,Brans:1961sx}.
    This theory, formally part of the Horndeski class, is strongly constrained in the solar system, e.g. by the Cassini flyby ~\cite{Bertotti:2003rm}. In order for the fifth force to be consistent with  these local experimental constraints, it must be suppressed close to  matter sources, e.g. through a non-linear ``screening'' mechanism~\cite{Vainshtein:1972sx,Babichev:2009ee,Babichev:2013usa,Joyce:2014kja}. In contrast to GR, whose coupling to matter and self-coupling are completely
    determined by
    unitarity and Poincar{\'e} invariance~\cite{Schwartz:2014sze},   
    massless scalars allow for different  screening mechanisms,  e.g. through  scalar  or  derivative self-interactions~\cite{Joyce:2014kja}. Examples of the former are chameleon ~\cite{Khoury:2003aq}
	 and symmetron~\cite{Hinterbichler:2010es} screening, while
	 examples of the latter include kinetic screening (or $k$-mouflage)~\cite{Babichev:2009ee} and Vainshtein screening~\cite{Vainshtein:1972sx,Babichev:2013usa}.
In more detail, kinetic screening is activated  when the first derivative of the scalar field  (i.e. the Newtonian acceleration) exceeds a certain threshold, while
the relevant quantity for Vainshtein screening is
 the second  (i.e. curvature) derivative of the scalar field.
 In this work, we focus on  kinetic screening, although there are significant similarities between these two derivative-based types of screening.
    
    A class of theories that admit kinetic screening is given by  $K(X)$ theories\footnote{These are usually referred to as  $P(X)$ theories in flat space, and as $K(X)$ theories in curved space.}~\cite{Joyce:2014kja}.
        These theories, also known as $k$-essence,
      are described by the action
	\begin{equation} \label{eq:action}
  	  S = \int d^4 x \sqrt{-g} \left[\frac{M^2_{\rm Pl}}{2} R + K(X) \right] + S_m(\Psi_i,\Tilde{g}_{\mu\nu}) \,,
	\end{equation}
	where $M_{\rm Pl}$ is the Planck mass, $X=g^{\mu \nu} \partial_\mu \varphi \partial_\nu \varphi$, $K(X)$ is a function that defines the theory, $S_m$ denotes the matter action, $\Psi_i$ represents a set of matter fields, and $\Tilde{g}_{\mu \nu}= \Phi^{-1}g_{\mu \nu} $, with the function $\Phi(\varphi)$ defining a conformal coupling to matter.    
    The simple polynomial
    \begin{eqnarray} \label{eq:K_intro}
        K=-\frac{1}{2}X+\frac{\beta}{4\Lambda^4} X^2-\frac{\gamma}{8\Lambda^8} X^3+\ldots  \,,
    \end{eqnarray}
    with $\Lambda$ an energy scale,
	 allows for screening to develop in spherical symmetry~\cite{Babichev:2009ee,Brax:2012jr,terHaar:2020xxb}
	 and axisymmetry~\cite{Kuntz:2019plo}, for certain
	 values of 	 $\beta < 0$ and $\gamma>0$. In the non-linear screening regime, when $X \gtrsim \Lambda^4$, one would expect the dynamics to be outside the EFT regime. Luckily, $K(X)$ theories are radiatively stable
	 in the non-linear regime,     and the scale $\Lambda$ should be interpreted as a strong-coupling scale and not as the cut-off of the theory~\cite{deRham:2014wfa, Brax:2016jjt} (see App.~\ref{app:eft} for details).  In addition, the conformal coupling of the scalar to matter is radiatively stable against corrections from the matter sector~\cite{Hui:2010dn}.

	 Although the speed of sound
	 in these
	 theories can be  superluminal~\cite{Aharonov:1969vu}, this does not imply the breakdown of causality~\cite{Armendariz-Picon:2005oog,Bruneton:2006gf,Babichev:2007dw}. Indeed, the initial-value (Cauchy) problem is well-posed for a large class of
	 $K(X)$ theories, with or without screening~\cite{Armendariz-Picon:2005oog,Bruneton:2006gf,Babichev:2007dw,Brax:2014gra,Bezares:2020wkn}.\footnote{This is a nice feature, although one should note that EFTs with perfectly healthy UV completions may suffer from a breakdown of hyperbolicity (and thus of the Cauchy problem) in the IR~\cite{Solomon:2017nlh,Allwright:2018rut,Lara:2021piy}. Thus, the requirement of hyperbolicity is not informative on the values of the Wilson coefficients, although it has practical importance for numerical relativity (see however~\cite{Cayuso:2017iqc,Allwright:2018rut,Lara:2021piy,Franchini:2022ukz}).} On the other hand,
	 positivity bounds~\cite{Adams:2006sv} may prevent the EFT
	 from having a standard local, unitary, Lorentz-invariant completion.
	 It is unclear whether this is true for all the choices of $K(X)$ that allow for screening~\cite{Chandrasekaran:2018qmx,Davis:2021oce, Aoki:2021ffc}, and whether allowing for Lorentz violations in the UV may allow for relaxing these bounds. (Note that kinetic-like screening can also be achieved by breaking Lorentz symmetry explicitly at low energies~\cite{Zlosnik:2006zu,Blanchet:2011wv,Bonetti:2015oda}). An alternative to the standard UV completion could be considered  by means of classicalization~\cite{Dvali:2010jz}.

$k$-essence and similar theories are relevant for the DE/DM problems described above. Self-acceleration of the universe can be obtained with models of the type $K(X,\varphi)$~\cite{Armendariz-Picon:2000nqq, 		Armendariz-Picon:2000ulo}, although those  generically do not lead to screening~\cite{Brax:2012jr}. Purely kinetic models $K(X)$ can also serve as DE, although with some degree of fine tuning~\cite{Scherrer:2004au,Giannakis:2005kr}.  Models that produce a  MOND-like phenomenology on galactic scales are also of this type, sometimes featuring complex scalar fields and additional non-derivative self-interactions~\cite{Khoury:2014tka,Berezhiani:2015bqa,Hertzberg:2021fvu}.  The class of $K(X)$ models can also generate ``anti-screening'', where the fifth force  increases near  matter~\cite{Hertzberg:2022bsb}. An interesting feature of these models is that they do not violate the positivity bounds, and therefore they are  expected to have a standard UV completion.
    
    In the non-linear regime, $k$-essence theories are non-trivial to work with.  Recently, significant effort has been directed at producing numerical simulations of the full non-linear dynamics  in  both $k$-essence  ~\cite{Babichev:2016hys,Bernard:2019fjb,Bezares:2020wkn,terHaar:2020xxb,Bezares:2021yek,Bezares:2021dma,Lara:2021piy,Shibata:2022gec}  and in theories with Vainshtein screening ~\cite{Dar:2018dra,Gerhardinger:2022bcw} (in the flat spacetime limit)\footnote{Simulations have also been performed in other scalar-tensor theories in the Hordenski class that do not exhibit screening~\cite{Kovacs:2020pns,Figueras:2020dzx,Figueras:2021abd,East:2022rqi,AresteSalo:2022hua,Ripley:2022cdh}.}.
    These breakthroughs allowed for studying stellar oscillations~\cite{Bezares:2021yek,Shibata:2022gec}, gravitational collapse~\cite{Bezares:2021yek} and neutron star mergers~\cite{Bezares:2021dma}. These simulations, as well as numerical results for the two-body problem in the stationary limit~\cite{Hiramatsu:2012xj,Kuntz:2019plo,White:2020xsq} (see also Ref.~\cite{Braden:2020zfa}), indicate that the phenomenology of screening in dynamical regimes and beyond spherical symmetry presents non-trivial differences from the static and spherically-symmetric case. This modified phenomenology may include a  partial breakdown of the screening mechanism~\cite{Bezares:2021yek,Bezares:2021dma} and allow for further constraints on the parameter space of these theories. In addition to the inherent difficulties of $k$-essence theories, a further obstacle in a fully numerical approach arises for the cosmologically motivated models where $\Lambda \sim (H_0 M_{\mathrm{Pl}})^{1/2}$,  as the latter
    implies a huge separation between the cosmological scales and the local scales relevant for the solar system or compact object binaries. It is thus important to understand in more detail, analytically as much as possible, the physics of kinetic screening.
 
    In this work, we will focus on the two-body problem in theories with kinetic screening. We consider this problem first analytically, providing a  decomposition of the scalar equation and solving it with various approximation techniques,
    and we  then numerically check their validity. In particular, we will discover a partial breakdown of the screening mechanism in the regime where one would expect it to operate. We will consider the two-body problem for different choices of the kinetic function $K(X)$, including a modification of DBI theory that allows for ``opposite" DBI screening~\cite{Burrage:2014uwa}, i.e.
    \begin{equation}
        K(X)= \Lambda^{4}\sqrt{1-\Lambda^{-4}X} \,.
    \end{equation}

    Phenomenologically, the two-body problem is relevant in several astrophysical scenarios, e.g. in the solar system and in binary pulsars, where
    tests of gravity have historically been performed~\cite{Taylor:1982zz,Bertotti:2003rm,Will:2014kxa,Berti:2015itd,Kramer:2021jcw,Fienga:2023ocw}, and more recently in the merging binaries of compact objects detected by gravitational wave (GW) experiments~\cite{Will:2014kxa,Berti:2015itd,LIGOScientific:2021sio}. While the
    calculations of this paper are only at leading post-Newtonian (PN) order and are
 clearly  inadequate to quantitatively describe binary mergers, they do nevertheless allow for a qualitative insight into the dynamics of binary systems beyond GR.

    This paper is organized as follows. In Sec.~\ref{sec:setup} we will provide the field equations of $k$-essence, show how they can be reformulated, using the  Hodge-Helmholtz decomposition, review  kinetic screening and illustrate it in the case of  isolated objects.  In Sec. \ref{sec:k2b2} we will describe the analytical approximations and  the numerical formalism 
    that we use to study the two-body problem in $k$-essence at leading PN order in the scalar sector. In Sec. \ref{sec:kbub} we will focus on a specific finding of our investigation - the appearance of pockets of  linear dynamics inside a region
that  would be in a non-linear regime in the absence of the second object.  Thus far,
our results will be either general or focused on a polynomial $k$-essence model. In Sec. \ref{sec:k-other} we will instead explore  other models, including  ``opposite" DBI screening and anti-screening. We will summarize our results in Sec. \ref{sec:fin}. Some details on the regime of validity of EFTs in theories with kinetic screening and on the regularization of point-particle divergences are  presented in Appendices \ref{app:eft} and \ref{app:coloumb}, respectively. In App.~\ref{app:gaba} we comment on the parallel between the Hodge-Helmholtz decomposition and classical dual reformulation of self-interacting theories developed in Refs.~\cite{Gabadadze:2012sm,Padilla:2012ry}. Finally, validations of our numerical code are described in App.~\ref{app:num_test}. Throughout this paper, we will employ a mostly plus metric  signature $(-+++)$ and natural units $c=\hbar=1$, with $M^2_{\rm Pl}=1/(8\pi G_\mathrm{N})$. Spatial components are denoted with Latin letters, vectors in $\mathbb{R}^3$ are in boldface and the unit vectors carry a hat.

	%%%%%%%%%%%%%%%%%%%%
	%%%%%%%%%%%%%%%%%%%
	\section{Setup} \label{sec:setup}
	%%%%%%%%%%%%%%%%%%%%%
	%%%%%%%%%%%%%%%%%%%%

    \subsection{$k$-essence equations of motion} \label{sec:cov}
    
	The action for a $k$-essence scalar-tensor theory is given by Eq.~\eqref{eq:action}.   Matter is  assumed to be minimally coupled to the conformal metric\footnote{One can also perform a field redefinition and work with the conformal metric directly~\cite{Hui:2009kc}. This is usually referred to as ``Jordan frame'', as opposed to the Einstein frame used in this paper.} $\Tilde{g}_{\mu \nu}= \Phi^{-1}g_{\mu \nu}, $ for whose conformal factor we consider the
	following expansion at leading order $\varphi/M_{\rm Pl}$:
  	\begin{eqnarray}
  	  \Phi^{-1} \approx 1+\frac{\alpha}{M_{\rm Pl}} \varphi\,.
	\end{eqnarray}
 	In spherical symmetry, screening is  a robust consequence of this action [for appropriate choices of $K(X)$], even when considering higher order corrections to this expansion~\cite{Lara:2022gof}.
    	From the action, the equations of motion are
	\begin{align}
  	  &G_{\mu \nu} = \frac{1}{M^2_{\rm Pl}} (T_{\mu \nu}+T^{\varphi}_{\mu \nu}) \,, \\
  	 & T_{\mu \nu} = \frac{2}{\sqrt{-g}}\frac{\delta S_m}{\delta g_{\mu \nu}} \,, \\
   &     T^{\varphi}_{\mu \nu} = K(X)g_{\mu \nu} - 2 K_{X} \partial_\mu \varphi\partial_\nu \varphi \,, \\
  &      \nabla_\mu (K_{X} \nabla^\mu \varphi) = \frac{1}{2}\frac{\alpha}{M_{\mathrm{Pl}}} T \,, \label{eq:KG}
	\end{align}
	where $G_{\mu \nu } = R_{\mu \nu} - R g_{\mu \nu}/2$ and $R_{\mu \nu}$ are the Einstein and Ricci tensors
	for the metric $g_{\mu\nu}$,
  	$T_{\mu \nu} $ (with $T=g_{\mu \nu}T^{\mu \nu}$) and $T^{\varphi}_{\mu \nu}$ are the matter and the scalar energy-momentum tensors, and $K_X \equiv \partial K/\partial X$.
    
    	Let us start by defining
	\begin{eqnarray} \label{eq:chi_4v}
  	  \chi_\mu \equiv K_{X} \nabla_\mu \varphi \,.
	\end{eqnarray}
	In the absence of matter sources, this vector is covariantly conserved and represents the Noether current associated with the shift symmetry $\varphi \to \varphi + c$. Let us then perform a Hodge-Helmholtz decomposition of this current into a longitudinal component $\partial_\mu \psi$ and a transverse component $B_\mu$:
	\begin{align}
  	  &\chi_\mu =- \frac{1}{2} \nabla_\mu \psi + B_\mu \,, \label{eq:helmh_4v}\\
  &      \nabla_\mu B^{\mu}=0 \,. \label{eq:divB_4v}
	\end{align}
	To check that this decomposition is unique and well-defined,
	one can compute the divergence $\nabla_\mu \chi^\mu=\Box \psi$, with $\Box=\nabla_\mu \nabla^\mu$. One can then conclude that $\psi$ is uniquely determined if the D'Alembertian is invertible, which is the case if $\psi$ is
	given appropriate initial and boundary conditions. $B_\mu$
	can then be determined unambiguously\footnote{More generally, the decomposition~\eqref{eq:helmh_4v} is a consequence of the Hodge decomposition theorem, which states that any $p$-form $\omega$ on a compact, Riemannian manifold can be uniquely decomposed as $\omega= d\Psi + d^\dag \beta + \gamma$, where $d^\dag$ denotes a codifferential and $\gamma$ is a harmonic form defined by $\Delta \gamma=0$ with $\Delta=(d+d^\dag)^2$~\cite{Nakahara:2003nw}. In coordinates, this statement leads to \eqref{eq:helmh_4v}, provided that the harmonic component $\gamma$ vanishes. This will indeed be the case for appropriate boundary/initial conditions. } from Eq.~\eqref{eq:helmh_4v}.
    
	By replacing the decomposition~\eqref{eq:helmh_4v} in Eq. \eqref{eq:KG}, one then gets the following 
 Klein-Gordon equation for the longitudinal mode:
	\begin{equation}\label{boxpsi}
  	  \Box \psi=-\frac{\alpha}{M_{\mathrm{Pl}}} T\,.
	\end{equation}
	As for $B_\mu$, taking a covariant derivative of Eq.~\eqref{eq:chi_4v} and anti-symmetrizing we obtain
	\begin{equation}
  	  \nabla_{[\nu} B_{\mu]} =K_{XX} \nabla_{[\nu} X \nabla_{\mu]}\varphi\,,
	\end{equation}
	where the antisymmetric part of a tensor $S_{\mu\nu}$ is defined as $S_{[\mu\nu]}=(S_{\mu\nu}-S_{\nu\mu})/2$. This equation can be put in manifestly hyperbolic form by taking a divergence [using also Eq.~\eqref{eq:divB_4v}], which leads to
	\begin{eqnarray} \label{eq:wave_B}
  	  \Box B^\mu - R^\mu_{\,\,\nu} B^\nu = J^\mu \,, \\
  	  J_\mu = 2\nabla^\nu[K_{XX} \nabla_{[\nu} X \nabla_{\mu]}\varphi] \,.
	\end{eqnarray}
	Note that this is formally the same as the equation for the relativistic vector potential in electromagnetism. 
 In particular, in FJBD theory $K_{XX}=0$,
 and this equation therefore implies  $B^{\mu}=0$ and $\varphi=\psi$. [This can also be seen directly from Eqs.~\eqref{eq:chi_4v}--\eqref{eq:helmh_4v}, recalling that $K_X=-1/2$ in FJBD theory]. Thus, we will refer to $\psi$ as the FJBD field.

By squaring Eq.~\eqref{eq:chi_4v}, one obtains $\chi^\mu \chi_\mu= K_X^2 X$. In order to express $X$ as a function of $\chi^\mu \chi_\mu$, $K_X^2 X$ needs to be a monotonic function of $X$. This therefore requires $1+2 X K_{XX}/K_X$ being sign definite. 
One can obtain the same condition by requiring invertibility of $\partial_\mu \varphi$ in terms of $\chi_\mu$ from the transformation \eqref{eq:chi_4v}. That  requires the Jacobian 
of Eq.~\eqref{eq:chi_4v}, i.e.
	\begin{equation} \label{eq:jacob}
 	\mathcal{J}_{\mu \nu} = K_X g_{\mu \nu} + 2 K_{XX} \partial_\mu \varphi \partial_\nu \varphi\,,   
	\end{equation}
to be sign-definite~\cite{Chandrasekaran:2018qmx}. Since $\rm{det\,}$$\mathcal{J}_{\mu \nu}\propto 1+2 X K_{XX}/K_X$,\footnote{This can be easily proven for generic $g_{\mu\nu}$ by projecting the Jacobian \eqref{eq:jacob} on a tetrad basis.} this yields again the same condition. By requring additionally invertibility 
for small values of $X$, one finally obtains the condition
	\begin{equation} \label{eq:X_cond_invert}
    	1+ \frac{2 K_{XX} X}{K_X}  > 0 \,.
	\end{equation}
	Remarkably, this  is the same condition that is found by requiring that the field equations of $k$-essence
    are strongly hyperbolic~\cite{Armendariz-Picon:2005oog,Bruneton:2006gf,Babichev:2007dw,Brax:2014gra,Bezares:2020wkn}.

	%%%%%%%%%%%%%%%%%
	\subsection{Non-relativistic and static limit} \label{sec:EoM_R3}
	%%%%%%%%%%%%%%%%%
    
	Let us now consider the scalar equation of motion
	at leading PN order, i.e. at leading order in $1/c$. To this purpose, let us note that if $c\neq 1$
	is reinstated, then $g_{\mu\nu}=\eta_{\mu\nu}+{\cal O}(1/c^2)$ and $\Box=\nabla^2+{\cal O}(1/c^2)$ (with $\nabla^2=\delta^{ij}\partial_i\partial_j$). The scalar equation \eqref{eq:KG} at leading PN (i.e. Newtonian) order is therefore simply
   	\begin{eqnarray} \label{eq:KG_NR}
  	  \partial_i (K_{X} \partial^i \varphi) = \frac{1}{2}\frac{\alpha}{M_{\mathrm{Pl}}} T \,.
	\end{eqnarray}
	For a binary system of point particles and again up to higher order corrections in $1/c$, one has
   	\begin{equation} \label{eq:T_point_2p}
  	  T=- m_a \delta^{(3)}(\bm{r}-\bm{r}_a(t))-m_b \delta^{(3)}(\bm{r}-\bm{r}_b(t))  \,,
	\end{equation}
	where $\bm{r}_{a,b}(t)$ are the two trajectories. It is then easy to check that if we find a solution $\varphi_{\rm static}(\bm{r},\bm{\bar{r}}_a,\bm{\bar{r}}_b)$ for the static problem (with the two particles at rest at positions $\bm{\bar{r}}_a,\bm{\bar{r}}_b$), the solution to Eq.~\eqref{eq:KG_NR} for two particles in motion with velocities $\ll c$
	can be obtained simply as
	$\varphi(t,\bm{r})=\varphi_{\rm static}(\bm{r},\bm{r}_a(t),\bm{r}_b(t))$. In the following we will therefore restrict, with no loss of generality (at least as long as one is working at Newtonian order), to the case of two static point particles.

   In terms of the (three-dimensional) Helmholtz decomposition\footnote{This decomposition was also introduced for $k$-essence in Ref.~\cite{Brax:2014gra}, although the solenoidal component was set to zero in spherical symmetry (see Sec. \ref{sec:an_sph}). In the context of MOND, the decomposition was introduced in Ref.~\cite{Milgrom:1986ib}, while the behavior of the two components was discussed for a particular type of $K(X)$ and in particular regions of space in a binary problem in Ref.~\cite{Bekenstein:2006fi} (see Sec. \ref{sec:kbub}).}     %
 %   [also by taking a static limit of Eq. \eqref{eq:helmh_4v}]
	%
	\begin{equation} \label{eq:helmh_3v}
  	  \bm{\chi}=-\frac{1}{2}\bm{\nabla} \psi + \bm{B}\,,
	\end{equation}
	 Eq.~\eqref{boxpsi} for $\psi$ therefore becomes the Poisson equation
	\begin{equation}
  	  \nabla^2 \psi =-\frac{\alpha}{M_{\mathrm{Pl}}} T \,.
	\end{equation}
	This equation can be solved for an $N$-body system simply by linear superposition. If one could assume $\bm{B}=0$, one would have to invert Eq. \eqref{eq:helmh_3v} in order to find $\varphi$, i.e., by squaring that equation, one would have to invert
        	\begin{equation} \label{eq:X_poly_eq}
  	  K_{X}^2X = \frac{1}{4}X_{\psi}\,,
	\end{equation}
with $X_\psi=(\nabla \psi)^2$. This is  possible if Eq.~\eqref{eq:X_cond_invert} is satisfied.
The implicit assumption  $\bm{B}=0$ was made, in the DBI case, in Ref.~\cite{Burrage:2014uwa}.
	However, it is not a priori clear that the solenoidal component $\bm{B}$ can be ignored. In the rest of the paper, we will discuss the role and  importance of this component.

	In $\mathbb{R}^3$ the decomposition
 that we are using, i.e. one  into  a longitudinal component (irrotational  vector field) and a transverse component (solenoidal vector field),   is further strengthened by the Helmholtz theorem~\cite{2017inel.book.....G}, which states that if all involved functions have appropriate asymptotic behavior, the decomposition \eqref{eq:helmh_3v} is unique and
	\begin{eqnarray}
  		  \psi &=& - \frac{1}{4\pi M_{\mathrm{Pl}}} \int d^3 \bm{r}' \frac{\alpha T(\bm{r}')}{|\bm{r}-\bm{r}'|} \,, \label{eq:Poisson_sol_psi}\\
   	  \bm{B} &=&  \bm{\nabla} \times \frac{1}{4\pi} \int d^3 \bm{r}' \frac{\bm{C}(\bm{r}')}{|\bm{r}-\bm{r}'|} \label{eq:B} \,, \\
  	  \bm{C}&\equiv& \bm{\nabla} \times \bm{\chi}\label{eq:C} \,.
	\end{eqnarray}
	From the definition of the vector $\bm{\chi}$
	[see Eq.~\eqref{eq:chi_4v}], one has
 \begin{eqnarray}
	 \bm{C} &=&  K_{XX} \bm{\nabla} X \times \bm{\nabla} \varphi\label{eq:C_source} \\
	 &=&  2 K_{XX} \epsilon_{ijk}  \partial_l \varphi \partial_j \partial_l \varphi  \partial_k \varphi \,,\nonumber
 \end{eqnarray}
 where $\epsilon_{ijk}$ is the totally antisymmetric Levi-Civita symbol. It is clear that the solenoidal component will be highly suppressed (or zero) in  highly symmetric regions/scenarios  where  $ \bm{\nabla} X$ and $\bm{\nabla} \varphi$ are parallel, or when non-linearities are suppressed.  
    
	Note that the total gravitational force
 (at leading PN order) 
 between  two bodies, separated by a distance $D$, is the sum of the Newtonian/GR force and the scalar fifth force. In FJBD theory the scalar force has a Newtonian-like behavior and just renormalizes the gravitational constant:
	\begin{eqnarray}
    	F_\mathrm{g} && = F_\mathrm{N}+F_\mathrm{FJBD} \\
    	&& = \left(G_N + \frac{\alpha^2}{4\pi M^2_\mathrm{Pl}}\right) \frac{m_a m_b}{D^2},
	\end{eqnarray}
	where the term in  brackets defines the effective gravitational constant. In theories with screening, the fifth force will exhibit a different behavior. In the following, we will ignore the usual Newtonian/GR component and focus on the scalar force.

	%%%%%%%%%%%%%%%%%%%%
	%%%%%%%%%%%%%%%%%%%
	\subsection{Isolated object} \label{sec:an_sph}
	%%%%%%%%%%%%%%%%%%%%
    
	Let us first briefly review the solution for an isolated object, extensively discussed elsewhere~\cite{Brax:2012jr,Brax:2014gra,Bloomfield:2014zfa,terHaar:2020xxb,Bezares:2021yek,Lara:2022gof}, from the perspective of the Helmholtz decomposition. In the case of a point particle or a spherically symmetric object, spherical symmetry implies that both $\bm{\nabla} \varphi$ and $\bm{\nabla} X$ must be parallel to the radial vector $\bm{r}$. Thus, from the discussion in Sec. \ref{sec:EoM_R3} [and particularly Eq.~\eqref{eq:C_source}], the solenoidal component must vanish. The solenoidal component will vanish also in other highly symmetric configurations, if there is only one vector in the problem that all  quantities need to be proportional to.
    
	Consider a point particle at the origin with mass $m$. For a quadratic choice of the kinetic function  of Eq.~\eqref{eq:K_intro}, i.e. 
	\begin{eqnarray} \label{eq:K_2}
	K(X)=K_2(X) \equiv -\frac{1}{2}X-\frac{1}{4\Lambda^4} X^2
	\end{eqnarray}
	the full solution to Eq.~\eqref{eq:KG_NR} can be expressed in terms of the generalized hypergeometric function as~\cite{Kuntz:2019plo}:
	\begin{eqnarray} \label{eq:1P_sol}
  	  \varphi\!\! &=&\!\! - \frac{1}{4 \pi r} m\frac{\alpha}{M_{\mathrm{Pl}}} \, \prescript{}{3}{F}_2 \Big[\frac{1}{4},\frac{1}{3},\frac{2}{3};\frac{5}{4},\frac{3}{2};-\Big(\frac{r_{\rm sc}}{r}\Big)^4\Big]\,, \label{eq:1p_sol} \\
  	  r_{\rm{sc}} \!\!&=&\!\!\frac{1}{\Lambda} \left( \frac{27}{4} \right)^{1/4} \sqrt{\frac{ m\alpha  }{4 \pi M_{\mathrm{Pl}}}}  \\
 	&=& 3 \times 10^{11} \mathrm{km} \times \nonumber \\
 	&&\left( \frac{\alpha}{0.1}\right)^{1/2} \left( \frac{\Lambda}{1.9 \times 10^{-3} \rm{eV}} \right)^{-1}   \left( \frac{m}{M_\odot}\right)^{1/2} \nonumber \,.
	\end{eqnarray}
	%
	%Sqrt[0.1]*c*sqrt[solar mass/(reduced planck mass)]/sqrt[hubble constant*reduced planck mass*c^2/hbar]*(27/4)^(1/4)/sqrt[4*Pi] to parsec
	The length scale that controls the solution is the kinetic screening radius $r_{\rm sc}$~\cite{Babichev:2009ee}. Fixing the value of $r_{\mathrm{sc}}$ determines the profile of $\varphi(r)/(m\alpha)$, although there is degeneracy among the individual parameters $\{m, \alpha, \Lambda \}$.
	One can compare this ``screened'' solution to the FJBD one [Eq.~\eqref{eq:Poisson_sol_psi}]. For a point particle, the latter diverges at the particle's location, while the scalar field $\varphi$, as a result of the non-linear term in the kinetic function $K(X)$, remains finite, i.e. at small radii one has
	\begin{eqnarray} \label{eq:1p_scr}
  	  \varphi  \approx - 3.7 \Lambda  \sqrt{\frac{ m\alpha  }{4 \pi M_{\mathrm{Pl}}}} + 3 \Lambda^{4/3} \left(\frac{ m\alpha  }{4 \pi M_{\mathrm{Pl}}} \right)^{1/3} r^{1/3} \,.
	\end{eqnarray}
	The screening therefore acts as a physical
	``UV regulator" for the field.
	Expanding the full result \eqref{eq:1p_sol}  around  $\Lambda \to \infty$ , one instead obtains
	\begin{equation} \label{eq:point_pt_pert_exp}
  	  \varphi = -\frac{ m\alpha  }{4 \pi M_{\mathrm{Pl}}}\frac{1}{r} +  \frac{1}{5} \left( \frac{ m\alpha  }{4 \pi M_{\mathrm{Pl}}}\right)^3 \frac{1}{\Lambda^4 r^5} + \mathcal{O}(\Lambda^{-8})\,,
	\end{equation}
	which demonstrates that  screening  is a non-perturbative effect appearing only in the regime $X/\Lambda^4\gg 1$, i.e. in order to recover it one needs to resum all the terms in the perturbative expansion~\cite{Davis:2021oce}.

    Note that although the scalar field is finite at the origin,  the scalar gradient still diverges. This is, however, simply 
 an artifact of the point particle approximation, i.e.
 the scalar gradient goes to zero at the origin for a spherical star~\cite{terHaar:2020xxb}.
	In the following, when solving for the scalar field in a two body system, we will therefore  have to
	resolve (or ``regularize'') the Dirac deltas
	in order to allow for a numerical treatment of the problem. We will do so by utilizing a Gaussian density model for each point particle, i.e.
	\begin{equation} \label{eq:source}
  	  T_i =-\frac{ m_i}{(\sqrt{2\pi}\sigma)^3}\exp{-\frac{(\bm{r}-\bm{r}_i)^2}{2\sigma^2} } \,,
	\end{equation}
	where $\sigma$ is the width of the Gaussian and $\bm{r}_i$ is the position of the particle.
	The solution of Eq.~\eqref{eq:KG_NR} with this source cannot be
	expressed in closed form, even for a single particle, although Eq.~\eqref{eq:X_poly_eq} provides a closed form expression for $\partial_r \varphi$ [see Eq. \eqref{eq:X_quad_invert} below]. This expression, however, needs to be integrated in order to find the scalar profile $\varphi(r)$ for a single point particle.
    
\begin{figure}
  	  \centering
  	  \includegraphics[width=0.48\textwidth]{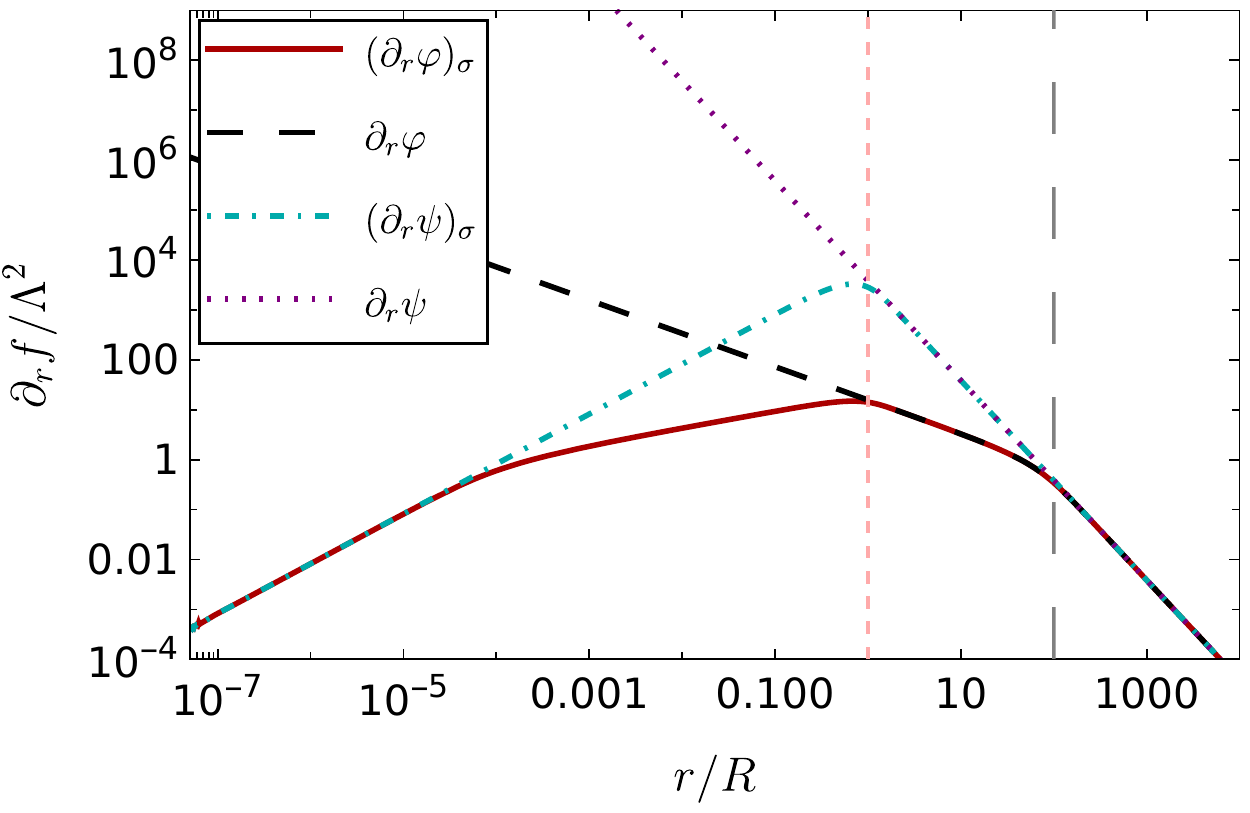}
  	  \caption{Scalar  gradients for the quadratic $k$-essence theory  described by Eq.~\eqref{eq:K_2} and for an isolated object located at $r=0$. The screening radius is $r_{\rm sc}=100 R $ (gray long dashed vertical line) and the effective radius of the Gaussian source  is $R=2\sigma$ (pink short dashed vertical line). $f=\{\varphi,\psi\}$ are respectively the $k$-essence scalar field and its irrotational component (which matches the FJBD gradient). Profiles are computed for a Gaussian source (index $\sigma$) and a Dirac delta (no index).}
  	  \label{fig:1pt_X}
	\end{figure}

	To test this regularization, we show in Fig. \ref{fig:1pt_X}  the scalar  gradient, calculated
	with a Dirac delta and a Gaussian source. As expected, the two profiles coincide outside the effective radius of the object ($R=2\sigma$).
	The same figure also shows the gradient of the FJBD field $\psi$, again for both sources.
	As expected from previous studies~\cite{Babichev:2009ee,terHaar:2020xxb,Bezares:2021yek}, the gradients of $\varphi$ (which can be physically interpreted as the fifth force) are suppressed with respect to the FJBD gradients $\partial_r \psi$, even inside the effective radius $R$.
 As the radial coordinate approaches the origin,  both gradients tend to zero for the regular Gaussian source, as dictated by spherical symmetry and regularity~\cite{terHaar:2020xxb,Bezares:2021yek}.
 Further details on  the Gaussian regularization will be presented in  App.~\ref{app:coloumb}.
    
%%%%%%%%%%%%%%%%%%%%
	%%%%%%%%%%%%%%%%%%%
	\section{The two-body problem: polynomial $k$-essence} \label{sec:k2b2}
	%%%%%%%%%%%%%%%%%%%%%
	%%%%%%%%%%%%%%%%%%%%
    
	Unlike the isolated object case of Sec. \ref{sec:an_sph}, the two-body problem is only axially symmetric, around the direction that connects the two particles.
  Let us define the coordinate system such that the particles (of masses $m_i$) are located on the $z$ axis at $z_i=\pm D/2$ , where $D$ is the separation and $i$ is an index running  on the two particles $a$ and $b$. We will work in cylindrical coordinates $(\rho,\vartheta,z)$, with $\rho=\sqrt{x^2+y^2}$ and $\vartheta=\arg(x,y)$.   In principle, the scalar field could depend on
  $\vartheta$, but because the source on the right-hand side of Eq.~\eqref{eq:KG_NR}
  does not, the dependence must be linear, i.e.
  $\varphi=L\vartheta+\bar{\varphi}(\rho,z)$ for some constant $L$. This would ensure that the left-hand side of Eq.~\eqref{eq:KG_NR} is independent of $\vartheta$. However, asymptotic flatness requires $\varphi$ approaching zero far from the two-body system, which in turn imposes $L=0$.

  Let us then solve the scalar equation \eqref{eq:KG_NR} with the source \eqref{eq:T_point_2p} and the polynomial
   kinetic function
 \begin{eqnarray} \label{eq:function_poly}
 	K(X) =\Lambda^4  \mathcal{K}_N(X) \,, \quad\mathcal{K}_N(X) =-  \sum^N_{n=1} \frac{1}{2n} \Big(\frac{X}{\Lambda^4} \Big)^n \,,
 \end{eqnarray}
 which allows for  screening as long as the leading order term has a negative coefficient~\cite{Davis:2021oce}. Note that the
 choice of the dimensionless series coefficient was made simply for computational convenience, and is not expected to qualitatively impact our results (see also Sec.  \ref{sec:kin_f_var}). In App. \ref{app:eft} we argue, based on Ref.~\cite{deRham:2014wfa}, that the non-linear regime of the above theory is in the EFT regime of validity for generic astrophysical scenarios and every $N > 1$.
 
From the analysis of the scalar profile around an isolated object in the previous section, we have seen that the screening starts  operating when $X \gtrsim \Lambda^{-4}$, and that the field strength is governed by  $\{m,\alpha,\Lambda\}$. For a generic polynomial function, in the region of deep screening the highest power of $X^N$ will dominate. For an isolated point particle, we then have, from Eq.  \eqref{eq:KG_NR},
\begin{eqnarray} \label{eq:KG_poly_N_sc}
\left( \frac{\partial_r\varphi}{\Lambda^2} \right)^{2N-1} \approx \left(\frac{ m \alpha  }{4 \pi M_{\mathrm{Pl}}\Lambda^2}\right) \frac{1}{r^2}\,.
\end{eqnarray}
Thus, the lengthscale that controls the scalar field profile for an isolated object  is parametrically the same as for  quadratic $k$-essence, i.e.
\begin{eqnarray} \label{eq:r_sc_N}
 r_\mathrm{sc} = c_N  \sqrt{\frac{ m \alpha  }{4 \pi M_{\mathrm{Pl}}\Lambda^2}} \,,
\end{eqnarray}
up to a numerical coefficient $c_N$ [for a quadratic kinetic function one has e.g. $c_2 =(27/4)^{1/4}]$. In particular, for $N=2$, from Eq. \eqref{eq:KG_poly_N_sc} one finds the small-radius expansion given by Eq.~\eqref{eq:1p_scr}, i.e. $\varphi \approx \mbox{const}+\, {\cal O}(r^{1/3})$.

Let us now turn to the binary problem and define $m_a \equiv m$ and $m_b \equiv m/q$, $q \geq 1$. Motivated by the previous discussion, we introduce the rescaled dimensionless variables
	\begin{eqnarray} \label{eq:tildas_units}
    	\mathsf{x}_i=\lambda x_i \,, \quad \kappa=\frac{ m \alpha  }{4 \pi M_{\mathrm{Pl}}} \left( \frac{\lambda}{\Lambda} \right)^2 \,, \quad \phi=\frac{\varphi \lambda}{\Lambda^2 } \,,\\
	\mathcal{X}=\frac{X}{\Lambda^4} \,, \quad  \Psi= \frac{\psi\lambda}{\Lambda^2 } \,, \quad \mathcal{B} = \frac{B}{\Lambda^2} \,, \quad \bm{\mathcal{C}}   = \frac{\bm{C}}{\lambda \Lambda^2(4\pi)} \,,
	\end{eqnarray}
	where the constant $\lambda$ (which has dimensions of a mass) is for the moment left free. Note that the
 rescaling of the Cartesian coordinates implies in particular the rescaling
$\varrho=\lambda \rho$. Moreover, from now on the spatial derivatives will be assumed to be taken with respect to the rescaled coordinates $\mathsf{x}_i$ unless otherwise specified.
  With these rescalings, the scalar equation \eqref{eq:KG_NR} takes the form
	\begin{eqnarray} \label{eq:KG_stationary}
   	\bm{\nabla} \cdot  \left(\bm{\nabla} \phi \sum^{N}_{n=1} \mathcal{X}^{n-1} \right) = \nonumber \\
    4 \pi\kappa  \left[ \delta^{(3)}\left(\bm{\mathsf{r}}- \frac{\mathsf{D}}{2}\bm{\hat{z}}\right)+\frac{1}{q}  \delta^{(3)}\left(\bm{\mathsf{r}}+ \frac{\mathsf{D}}{2}\bm{\hat{z}}\right)
    \right]
	\end{eqnarray}
    We will from now on  use the  scale invariance of Eq. \eqref{eq:KG_stationary} to set $\mathsf{D}=1$, and thus $\lambda=D^{-1}$, without loss of generality.  Thus,  the parameter space of the problem is defined by the (square of the) ratio of the screening radius of the more massive object and the inter-particle separation,
 \begin{equation}
      \kappa = \frac{ m \alpha }{4 \pi M_{\mathrm{Pl}}\Lambda^2} \frac{1}{D^2} \propto \left(\frac{r_{\mathrm{sc}}}{D}\right)^2 \,,
 \end{equation}
and by the mass ratio $q$.
As mentioned earlier, to solve the two-body problem numerically we will need a finite representation for the Dirac deltas, i.e.
 Eq. \eqref{eq:source} for both sources, centered at $\mathsf{z}=\pm 1/2$.
	The appearance of a resolution length scale $\sigma$ now extends
	the dimension of the parameter space from two to three:
	$\kappa$, $q$ and $\mathsf{R}$, where $\mathsf{R}=R/D=(2\sigma)/D$.
    
	In the following, for concrete  calculations we will focus on a quadratic model ($N=2$), which makes
	Eq.~\eqref{eq:X_poly_eq} solvable analytically. However, we will also provide analytic arguments supporting
 the (qualitative) applicability of
 these results to more general polynomial functions.

    %%%%%%%%%%%%%%%%%%%%
    %%%%%%%%%%%%%%%%%%%
    \subsection{To \textbf{\textit{B}} or not to \textbf{\textit{B}}} \label{sec:analy}
    %%%%%%%%%%%%%%%%%%%%%
    %%%%%%%%%%%%%%%%%%%%

    As discussed in Sec. \ref{sec:an_sph}, a perturbative treatment of the dynamics is only useful outside the screening region(s). In that regime, the solenoidal component is suppressed relative to the irrotational one, as $\mathcal{K}''\ll 1$ [see Eq.  \eqref{eq:C_source}]. Inside the screening region(s), the equation of motion is highly non-linear, and an exact solution cannot be found for the two-body problem.
    We will therefore attempt
    two approximations, whose validity we will evaluate by self-consistency and
    by comparing to our fully numerical results\footnote{One can perform a field rescaling ~\cite{McManus:2017itv,Renevey:2020tvr,Renevey:2021tcz} or introduce an auxiliary field at the level of action (dual formulation) that replaces the self-interacting terms~\cite{Gabadadze:2012sm,Padilla:2012ry} in order to obtain a well-defined perturbative expansion in the screening region.  In App.~\ref{app:gaba} we show that the dual formulation is in fact equivalent to the Helmholtz decomposition. We have not been able to solve the equation for the solenoidal component \eqref{eq:wave_B} in a binary problem even after such a reformulation. Other analytic approaches,  based e.g. on perturbing the scalar field around the background field generated by a fictitious isolated body located at the center of mass of the system~~\cite{Andrews:2013qva}, or on a sort of effective one-body approach~\cite{Kuntz:2019plo}, have also failed to solve the two-body problem in a controlled perturbative manner in theories with kinetic/Vainshtein screening.  }.

As the Helmholtz decomposition \eqref{eq:helmh_3v} breaks the problem into a straightforward part (irrotational component) and a complicated one (solenoidal component), most of our focus will be on understanding the solenoidal component. Let us then outline its vectorial structure.   As elaborated in the introduction of this Section, from axial symmetry and asymptotic flatness one has $\phi=\phi(\varrho,\mathsf{z})$, and thus from the Helmholtz decomposition \eqref{eq:helmh_3v}
one can conclude that $\mathcal{B}^\vartheta=0$. As the source $\bm{\mathcal{C}}$
is proportional to $\bm{\nabla} \phi \cross \bm{\nabla} \mathcal{X}$, and these two vectors are spanned by $\{\hat{\bm{\varrho}},\hat{\bm{\mathsf{z}}}\}$, the only non-zero component of the source vector will be $\mathcal{C}^\vartheta$. From Eq. \eqref{eq:B} it then follows%
\begin{eqnarray}
   		  \mathcal{B}_{\varrho} &=&  - \int d  \mathsf{V}' \partial_{\mathsf{z} } \left( \frac{1}{|\bm{\mathsf{r}}-\bm{\mathsf{r}}'|} \right) \mathcal{C}^\vartheta(\varrho',\mathsf{z}') \,, \label{eq:B_cyl_comp} \\
     		  \mathcal{B}_\mathsf{z} &=&   \int d\mathsf{V}' \frac{1}{\rho}\partial_{\varrho } \left( \frac{\varrho}{|\bm{\mathsf{r}}-\bm{\mathsf{r}}'|} \right) \mathcal{C}^\vartheta(\varrho',\mathsf{z}') \,.  \nonumber
\end{eqnarray}
It is clear that these two components satisfy the zero-divergence condition \eqref{eq:divB_4v}
\begin{eqnarray}
\frac{1}{\varrho}\frac{\partial(\varrho\mathcal{B}_\varrho)}{\partial \varrho} = - \frac{\partial \mathcal{B}_\mathsf{z}}{\partial \mathsf{z}} \,.
\end{eqnarray}
If the source $\mathcal{C}^\vartheta$ is non-zero, as generically expected, the solenoidal component is expected to be non-vanishing, unlike in the example given in Sec. \ref{sec:an_sph}.

	\subsubsection{Linear superposition approximation}

	A very simple approximation is  to consider the \textit{linear superposition} of the \textit{full} single particle  solutions  [given by Eq.~\eqref{eq:1P_sol} for the quadratic kinetic function], i.e.
	\begin{eqnarray} \label{eq:linear_sp}
  	  \phi = \phi_a + \phi_b \,,
	\end{eqnarray}
	where $ \phi_{a,b}$ are the two solutions.
	We expect this ansatz to work well when the screening regions of the individual bodies do not overlap, i.e. in the limit $\kappa \ll 1$. In that situation, non-linearities are strong only in the vicinity of each  body, and are sourced by the body itself (in isolation). These non-linearities are therefore already captured by  Eq.~\eqref{eq:1P_sol}.
	This approximation will however
receive non-trivial and a priori uncontrolled corrections when the screening regions of the individual bodies  overlap.
    
 Note that even when $\kappa \ll 1$, the solenoidal component $\bm{\mathcal{B}}$ will be non-zero. In this regime, we can just insert the ansatz \eqref{eq:linear_sp} into Eq. \eqref{eq:helmh_3v}   and get a (complicated) expression for $\mathcal{B}^2=\bm{\mathcal{B}} \cdot \bm{\mathcal{B}}$. The latter can be simplified on the plane $\varrho=0$, where both the solenoidal and irrotational components present maximal amplitudes:
    \begin{eqnarray}
        \mathcal{B}^2(\varrho=0,z) &=& \frac{64}{3}f(a)^2 f(b)^2 \big[s(b)f(a)+s(a)f(b) \big]^2 \,, \nonumber\\
        f(i) &\equiv& \sinh{\Big[\frac{1}{3}\arcsinh{\left(\frac{\mathsf{r}_{\mathrm{sc},i}}{\mathsf{z}-\mathsf{z}_i}\right)^2} \Big]} \,,\\
        s(i) &\equiv& \mathrm{sgn}(\mathsf{z}-\mathsf{z}_i) \,.
    \end{eqnarray}
 In the regime where the superposition approximation is valid $(\kappa \ll 1)$, we find that the irrotational component dominates upon the solenoidal one, as the dimensionless ratio between their kinetic energies is
 \begin{eqnarray}
	 \frac{\mathcal{X}_\Psi(0,\mathsf{z})}{\mathcal{B}^2(0,\mathsf{z})} \sim \frac{81 q^2}{16 \kappa^4} \left(1-2 \mathsf{z} \right)  \,,
 \end{eqnarray}
 when $\kappa \to 0$.
Note that this is a non-trivial result, as it is valid not only in the perturbative regime (outside the screening regions of the individual bodies), but also in the screening region of each body.
Indeed, near the body positions $\mathsf{z} = \pm 1/2$, one recovers (by construction) the isolated object solutions, at leading order. 
Similarly, the isolated object solution can be recovered in the extreme mass ratio limit. Indeed, expanding for $q\gg 1$ we find
 \begin{eqnarray}
	 \mathcal{B}^2(0,\mathsf{z}) \sim \mathcal{O}(q^{-2}) \,,\quad \mathcal{X}_\Psi(0,\mathsf{z})  \sim \mathcal{O}(q^{0}) \,,
 \end{eqnarray}
 which shows that the solenoidal component is suppressed, as expected.

 \subsubsection{Irrotational approximation}

Motivated by the previous discussion, one can  start from the Helmholtz decomposition \eqref{eq:helmh_3v},  ignore the solenoidal component and invert Eq.~\eqref{eq:X_poly_eq} to find $X$. We will refer to this approach as the \textit{irrotational (or longitudinal) approximation}. Let us consider first a general kinetic function $K(X)$, and emphasize that the kinetic energy $X$ obtained in this way is an infinite series in $\Lambda^{-2}$ (although this scale is absorbed in the parameter $\kappa$). Indeed, in the case of an isolated object or other highly symmetric configurations, this result matches  the full result, as shown in Sec. \ref{sec:an_sph}. Once  found $X$, one can reconstruct the field by integrating Eq.~\eqref{eq:helmh_3v} (with $\bm{B}=0$), i.e.
     \begin{eqnarray} \label{eq:phi_from_psi}
\phi(\varrho,\mathsf{z})=-\frac{1}{2}\int^\varrho_\infty d\tilde{\varrho} \frac{\partial_{\tilde{\varrho}} \Psi}{\mathcal{K}'(\mathcal{X})}  \,.
 \end{eqnarray}
(Note that the general solution would include an additional arbitrary function of $\mathsf{z}$, which is however forbidden by requiring that $\phi$ vanishes far from the source.) For example, for the quadratic function of Eq.~\eqref{eq:K_2},  one  finds the following closed form for $\mathcal{X}$: 
    \begin{eqnarray} \label{eq:X_quad_invert}
\mathcal{X}&=&\frac{1}{3} \left(Y^{1/3}+Y^{-1/3}-2 \right) \,, \\
Y&=&\frac{3 \sqrt{3} \, (27 \mathcal{X}_\Psi^2+4 \mathcal{X}_\Psi)^{1/2}+27 \mathcal{X}_\Psi+2}{2} \,. \nonumber
    \end{eqnarray}
[from Eq. \eqref{eq:X_poly_eq}; see Fig. \ref{fig:X_Xpsi}].

Once determined $\mathcal{X}$ and $\phi$
 in this irrotational approximation, one
 can check the validity of the latter by computing the solenoidal component from Eq.~\eqref{eq:C_source}. In this way, one could devise an iterating scheme in order to solve the problem self-consistently. We will however try to understand if there are regimes where the solenoidal component is parametrically suppressed and the irrotational approximation  is valid to leading order. We are thus interested in comparing $\mathcal{B}$ with $-\bm{\nabla} \Psi/2$. As the same derivative power will act on the radial distance in the denominator on the right-hand side of Eq. \eqref{eq:Poisson_sol_psi} as in Eq. \eqref{eq:B_cyl_comp},  it follows that if the source of the solenoidal component  $\mathcal{C} \equiv |\bm{\mathcal{C}} |=\mathcal{C}^\theta$ is parametrically suppressed with respect to the irrotational one, i.e.
    \begin{eqnarray}
    S_\Psi=-\frac{1}{2}\kappa   \left[ \delta^{(3)}\left(\bm{\mathsf{r}}- \frac{1}{2}\bm{\hat{z}}\right)+\frac{1}{q}  \delta^{(3)}\left(\bm{\mathsf{r}}+ \frac{1}{2}\bm{\hat{z}}\right)
    \right]\,, \label{eq:S_psi}
	 \end{eqnarray}
 this will be also true for the magnitudes of the components themselves.

  The magnitude of the source of the solenoidal component in the irrotational approximation is given by
  	\begin{eqnarray}
\mathcal{C} &\approx& \mathcal{N}_K G_{\bm{\nabla}} \,,\\ \mathcal{N}_K &=&-  \frac{1}{8\pi}\frac{\mathcal{K}_{\mathcal{X}\mathcal{X}}}{\mathcal{K}_\mathcal{X}} \frac{d\mathcal{X}}{d\mathcal{X}_\Psi} |\bm{\nabla}\mathcal{X}_\Psi| \sqrt{\mathcal{X}_\Psi} \,, \label{eq:Fdel} \\
G_{\bm{\nabla}}&=& \sqrt{ 1-  \frac{(\bm{\nabla} \mathcal{X}_\Psi \cdot \bm{\nabla} \Psi)^2}{(\bm{\nabla} \mathcal{X}_\Psi)^2 \mathcal{X}_\Psi}} \,,\label{Gdef}
 \end{eqnarray}
 where the kinetic function and its derivatives are  functions of $\mathcal{X}=\mathcal{X}(\mathcal{X}_\Psi)$.
 As can be seen, the source $\mathcal{C}$ depends
 on the non-linear terms of the function (encoded in $\mathcal{N}_K$) and on the ``misalignment'' between $\bm{\nabla} \mathcal{X}\propto \bm{\nabla}\mathcal{X}_\Psi$ and
 $\bm{\nabla} \phi\propto \bm{\nabla}\Psi$ (encoded in $G_{\bm{\nabla}}$).
In order for the solenoidal component
to be significant, there must be an overlap between the supports of $\mathcal{N}_K$ and $G_{\bm{\nabla}}$.

\begin{figure}
        \centering
        \includegraphics[width=0.48\textwidth]{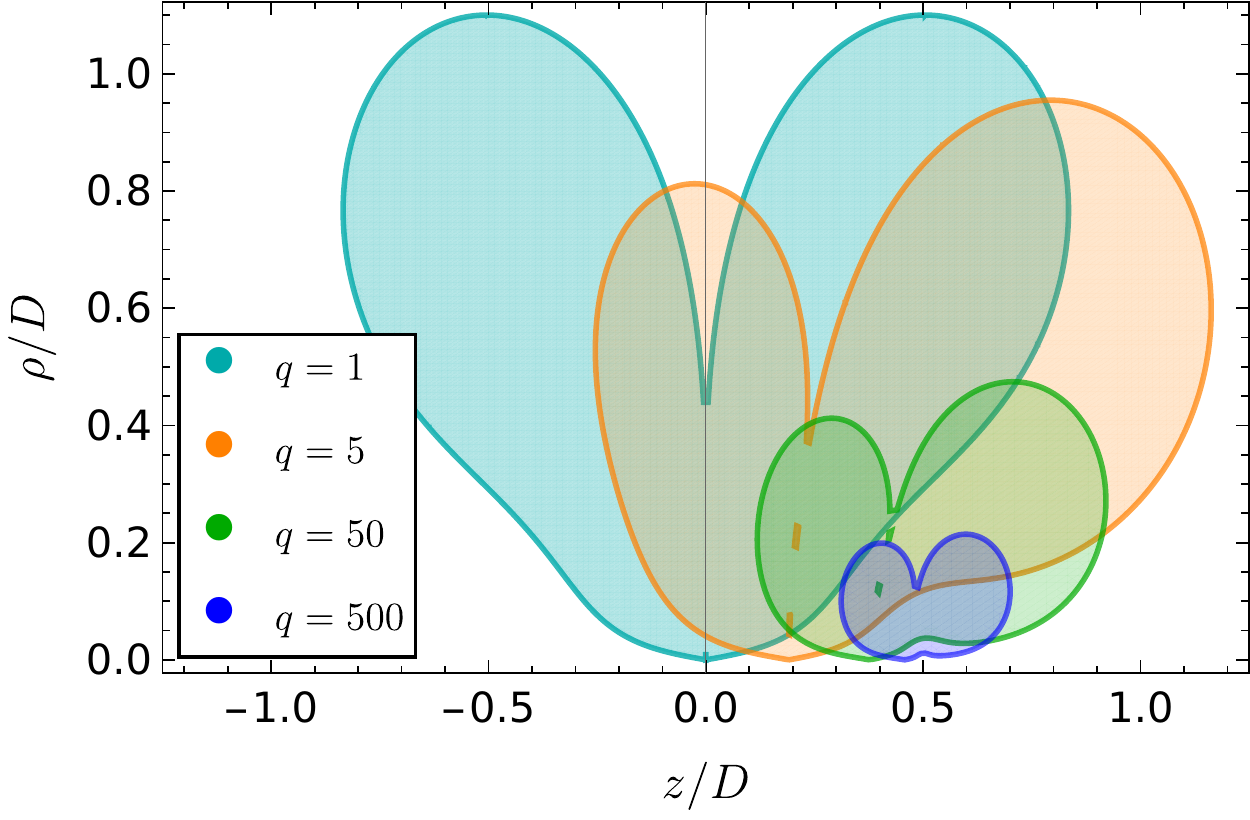}
        \caption{
        Support of the function $G_{\bm{\nabla}}$, which encodes the geometry of the source \eqref{Gdef},  defined by $G_{\bm{\nabla}}>0.1$, for four choices of the mass ratio $q=m_a/m_b$. The cylindrical coordinates $(\rho,z)$ are rescaled by the inter-particle separation, and the particles are located at $z_i=\pm D/2$.}
        \label{fig:dot_prod}
    \end{figure}

 Let us first consider  $G_{\bm{\nabla}}$.
 Since it depends only on the FJBD fields $\Psi$ and $\mathcal{X}_\Psi$, this quantity is independent of the screening radius and only depends on the mass ratio
$q$. From the definition \eqref{Gdef}, it is also clear that
$0  \leq G_{\bm{\nabla}} \leq 1$, with
 $G_{\bm{\nabla}} = 1$ when
 $\bm{\nabla} \mathcal{X}_\Psi$ and $\bm{\nabla} \Psi $ are orthogonal.
 In  Fig. \ref{fig:dot_prod}, we plot the support
 of $G_{\bm{\nabla}}$, defined as the region where $G_{\bm{\nabla}} \geq 0.1$, for various mass ratios.
  In the equal mass case, the support is symmetric around the center of mass of the system.
 As $q$ increases, the support  shrinks
 and gets shifted towards the smaller object (located at $\mathsf{z}_b = 1/2$).  In the limit $q \to \infty$ we find that $G_{\bm{\nabla}} \to 0$, i.e.  we recover the spherically symmetric solution and the solenoidal component  vanishes, as expected.

 The prefactor $\mathcal{N}_K$ has instead support in the screening region, where the non-linearities dominate and which is centered on the objects themselves, encoded in $\kappa\propto (r_{\rm sc}/D)^2$ and $q$. Let us first demonstrate two cases where the overlap between $\mathcal{N}_K$ and $G_{\bm{\nabla}}$
 is small. Consider first the equal mass limit $q \approx 1$ and $\kappa \ll 1$: the support of $\mathcal{N}_K$ shrinks
 (because the screening radii of the two objects shrink)
 and  the small overlap with the support of $G_{\bm{\nabla}}$ suppresses the source of the solenoidal component. When instead $q \gg 1$,  the support of $\mathcal{N}_K$ is mostly around the more massive body $a$, while  the support of $G_{\bm{\nabla}}$ is closer to the lighter body $b$ (see Fig. \ref{fig:dot_prod}), resulting again in a small overlap between $\mathcal{N}_K$ and $G_{\bm{\nabla}}$ and thus in a small source magnitude $\mathcal{C}$.
 Note that both of these cases ($q\approx 1$ and $\kappa \ll 1$; $q \gg 1$) are consistent with the intuition from the linear superposition approximation (for the quadratic $k$-essence), but the arguments presented here extend their validity to a generic kinetic function.

To gain some insight on the
remaining case ($\kappa \gg 1$ and  $q \approx 1$), let us specialize to
  the polynomial form \eqref{eq:function_poly} for the kinetic function.
 In the deep screening regime $\kappa \gg 1$, the highest power in the series dominates, and from \eqref{eq:X_poly_eq} one therefore concludes that
\begin{eqnarray} \label{eq:X_X_psi}
    \mathcal{X} \approx \mathcal{X}_\Psi^{1/(2N-1)}\,,
\end{eqnarray}
and consequently
\begin{eqnarray} \label{eq:F_k_deep_scr}
	\mathcal{N}_K &\approx& - \frac{\kappa}{8\pi} \frac{N-1}{(2N-1)} \frac{|\bm{\nabla} \hat{\mathcal{X}}_\psi|}{\sqrt{\hat{\mathcal{X}}_\Psi}} \,, \\
 \mathcal{X}_\Psi &\equiv& \kappa^2 \hat{\mathcal{X}}_\Psi(q) \,.
\end{eqnarray}
For  $N=1$,
 one recovers the FJBD theory result ($\mathcal{X}=\mathcal{X}_\Psi$ and $\mathcal{N}_K=0$).   Since $G_{\bm{\nabla}}$ does not depend on $\kappa$, one can therefore conclude that $\mathcal{C}\approx \mathcal{N}_K G_{\bm{\nabla}}\propto \kappa$.
 
\begin{figure}
        \centering
        \includegraphics[width=0.48\textwidth]{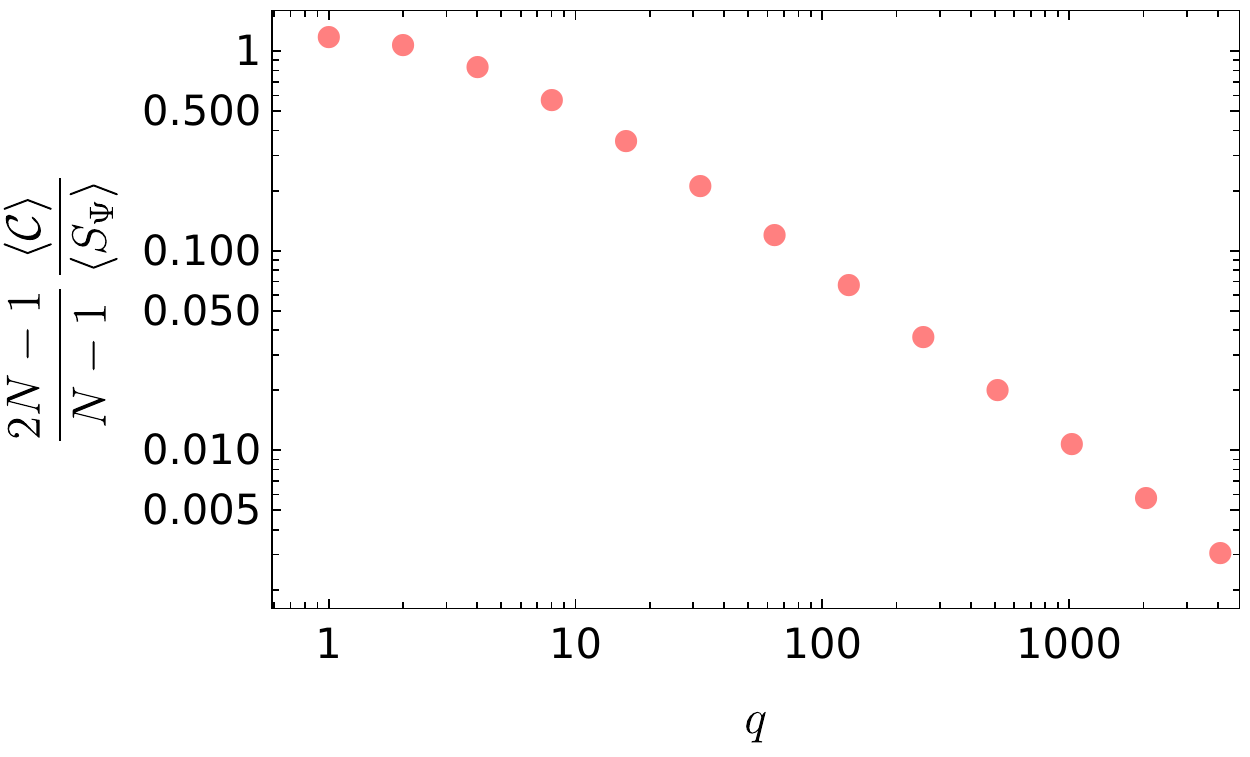}
        \caption{The ratio between the volume-averaged source of the irrotational component in the Helmholtz decomposition  \eqref{eq:helmh_3v}, $\langle S_\Psi \rangle$, and the (averaged) source of the solenoidal component, $\langle C \rangle $, as function  of the mass ratio $q$ in the deep screening regime $\kappa \gg 1$. The ratio is  multiplied by the factor $(2N-1)/(N-1)$, which depends on the choice of the kinetic function (in  polynomial form). 
        }
        \label{fig:source_q}
    \end{figure}

 Let us now compare the source of irrotational and the solenoidal components. As both $S_\Psi$ [see Eq. \eqref{eq:S_psi}] and $\mathcal{C}$ scale linearly with $\kappa$, their ratio will depend only on the mass ratio $q$.  Since $-\bm{\nabla} \Psi/2$ and $\bm{\mathcal{B}}$ [Eqs. \eqref{eq:Poisson_sol_psi}, \eqref{eq:B_cyl_comp}] depend on the volume integrals of their respective sources and both sources have a compact support, let us then compute the spatial averages of $\mathcal{C}$ and $ S_
\Psi$.  For the irrotational component, 
Eq. \eqref{eq:S_psi} yields
the average
 $\langle S_\Psi\rangle = $ $- \kappa (1+q^{-1})/(2V)$,
 where $V$ is the volume (larger than the individual supports of $\mathcal{C}$ and $ S_
\Psi$) over which the average is peformed.
  For $\mathcal{C}$, we calculated the average numerically for a set of values of $q$. In Fig. \ref{fig:source_q}, we show the ratio $\langle \mathcal{C} \rangle /\langle S_\Psi \rangle $,  multiplied by
 $(2N-1)/(N-1)$ to eliminate the dependence on $N$ [see Eq.~\eqref{eq:F_k_deep_scr}].
 For instance, for $q=1$ and $N=2$,
 the  ratio is $\langle \mathcal{C} \rangle /\langle S_\Psi \rangle  \approx 0.38$
 (Also note that the ratio is independent of the volume $V$). 
  These results demonstrate that the  solenoidal component is always suppressed with respect to the irrotational one even in the deep screening regime, although when $q \approx 1$ this suppression is less evident.

In conclusion, the analytic arguments of this section indicate that the  solenoidal component will be significantly suppressed  relative to the irrotational one when $\kappa \ll 1$ and/or $q \gg 1$ (for a generic kinetic function).
The suppression also holds in the deep screening regime $\kappa\gg1$ (for a polynomial kinetic function), although it becomes less pronounced for comparable masses ($q\approx 1$).

    %%%%%%%%%%%%%%%%%%%%
    %%%%%%%%%%%%%%%%%%%
    \subsection{Numerical solution} \label{sec:numerics}
    %%%%%%%%%%%%%%%%%%%%%
    %%%%%%%%%%%%%%%%%%%%

    %%%%%%%%%%%%%%%%%%%
    \subsubsection{Formulation} \label{sec:num_form}
   	 %%%%%%%%%%%%%%%%%%%%
	 
In order to validate the analytic approximations of Sec. \ref{sec:analy} and to understand the full behavior of the two-body dynamics, we have performed  numerical simulations for the case of a quadratic kinetic function\footnote{The same problem was also  studied numerically in Ref.~\cite{Kuntz:2019plo}. However, that analysis focused only on the binary's energy. Here we discuss several other aspects, and in particular the behavior of the scalar gradients. In addition, we use a different numerical method and provide a more refined code validation (see App. \ref{app:num_test}).}. The scalar equation of motion \eqref{eq:KG_stationary}, together with appropriate boundary conditions (to be described below), defines an elliptic boundary value problem. We have represented Eq. \eqref{eq:KG_stationary} in  cylindrical coordinates, and we have discretized it using a second-order finite difference scheme, regularizing the Dirac deltas with the Gaussian source of Eq.~\eqref{eq:source}.
 
 Our integration domain is the rectangle $[0,\varrho_{\rm out}] \times [-\mathsf{z}_{\rm out},\mathsf{z}_{\rm out}]$, 
 where the boundary values $\varrho_{\rm out}$ and $\mathsf{z}_{\rm out}$ are chosen to be larger than or at most comparable to the screening radii of the constituent object.
On the $\varrho =0$ plane, regularity requires  the boundary condition $\partial_\varrho \phi |_{\varrho \to 0}=0$. Defining our grid as $\varrho=i \mathsf{h}$ and $\mathsf{z}=j \mathsf{h}_\mathsf{z}$
(with $\mathsf{h}$ and $\mathsf{h}_{\mathsf{z}}$ the grid steps), regularity is then implemented by introducing the ghost point
$i=-1$ and taking $\phi(-1,j)=\phi(1,j)$. In order to regularize the coordinate singularity of Eq. \eqref{eq:KG_stationary} at $\varrho=0$, we apply the L'H\^{o}pital rule
\begin{eqnarray}
 \lim_{\varrho \to 0} \frac{1}{\varrho} \frac{\partial \phi}{\partial \varrho}=\frac{\partial^2 \phi}{\partial \varrho^2}   \,
\end{eqnarray}
to modify the scalar equation at $i=0$~\cite{MazumderPDE}.

On the other sides of the grid,  
we have used  two implementations of   Dirichlet boundary conditions. First, we have used the  superposition of the FJBD potentials  to set the scalar field on the boundary,
as long as the latter is sufficiently far from the screening radius. Moreover, after establishing that Eq.~\eqref{eq:phi_from_psi} provides a good approximation away from the objects, but inside the screening region, we have used it 
to set the 
 boundary scalar field in the case when the size of the domain is comparable to the screening radius, in order to reduce the size of the grid for the highly non-linear cases.

    After discretization, Eq. \eqref{eq:KG_stationary} yields the non-linear system
    \begin{eqnarray} \label{eq:G_num}
   G_{ij}[\{\phi(i',j')\}_{i'=i-1,i,i+1,j'=j-1,j,j+1}]=0  \,,
    \end{eqnarray}
    where $G_{ij}$ is a (non-linear) function of the discretized field at the neighboring points $i'$ and $j'$. We solve this non-linear system by using a Newton-Raphson method and an LU decomposition to compute the inverse of the Jacobian. We have set a tolerance of 
    $10^{-5}$ on the scalar field profile. In all runs grid size was several times smaller than the effective radius of the Gaussian source. 
    As initial guess for scalar profile in the Newton-Raphson method, we adopt several  choices including the superposition of FJBD potentials \eqref{eq:lin_trial} and the single particle solutions (Sec. \ref{sec:analy}). However, as the non-linearities become more important, corresponding to the growth of $\kappa$ in our units,  we  use Eq.~\eqref{eq:phi_from_psi} as initial guess. Moreover, to speed up the calculation we occasionally replace the Newton-Raphson method iteration with one computed with a Broyden method~\cite{Press:1992zz}.

    A description of our code's validation, including a comparison with the semi-analytic solution in the single particle case and  convergence tests, is left for App. \ref{app:num_test}.

    %%%%%%%%%%%%%%%%%%%%
    %%%%%%%%%%%%%%%%%%%
    \subsubsection{Results} \label{sec:results}
    %%%%%%%%%%%%%%%%%%%%%
    %%%%%%%%%%%%%%%%%%%%

\begin{figure*}[th]
\begin{tabular}{cc}
\includegraphics[width=.48\textwidth]{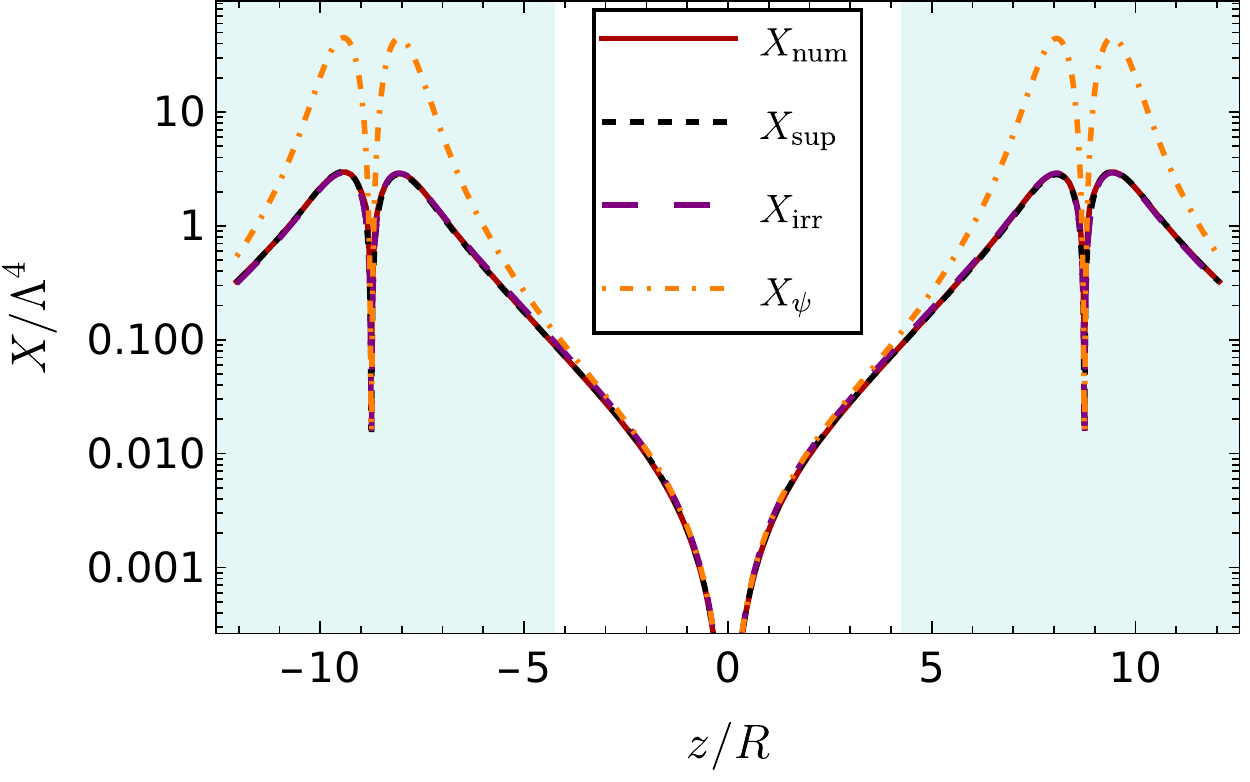}
\qquad
\includegraphics[width=.48\textwidth]{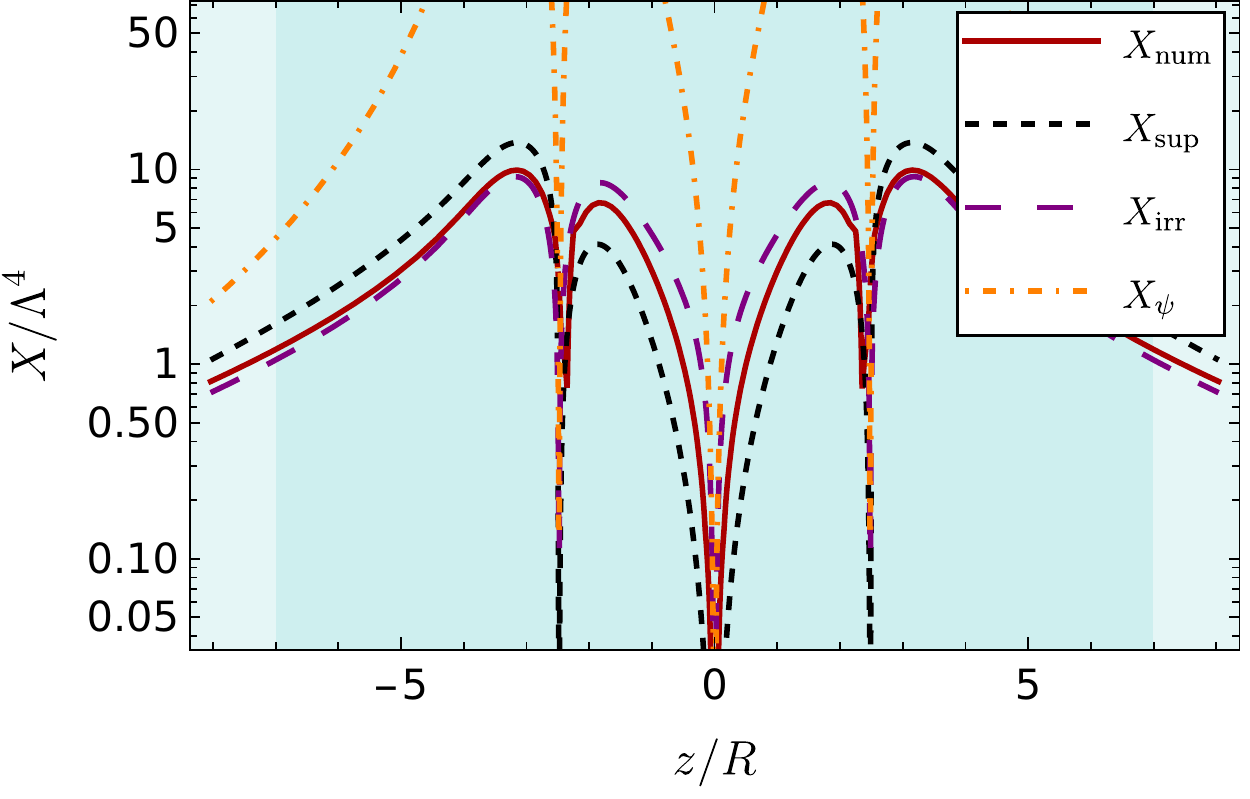}
\end{tabular}
\caption{Scalar kinetic energy on the plane $\rho=0$ for quadratic $k$-essence, calculated from our numerical results ($X_{\mathrm{num}}$, red solid line),
with the linear superposition approximation
($X_{\mathrm{sup}}$, black dashed line), and with the irrotational approximation ($X_{\mathrm{irr}}$, purple long-dashed line). Also shown for comparison is the FJBD result 
($X_{\psi}$, orange dot-dashed line).
Two equal-mass $(q=1)$ binary systems are considered:
 $r_{\rm sc}=4.5 R $, $D=17.5R$ (left) and $r_{\rm sc}=9.5 R $, $D=5R$ (right), with the origin placed at the geometric center.  The cyan shaded areas represent the individual screening regions of each body in isolation (ignoring the descreening  in the vicinity of the object's center), and the darker shade in the right panel denotes the overlap of these individual screening regions.}
	\label{fig:2pt_X}
\end{figure*}

    Consider first an equal-mass system 
    where   the screening regions of the constituent objects  do not overlap, i.e. $\kappa <1$. In Fig. \ref{fig:2pt_X} (left), we  show 
    the numerical solution for the scalar's kinetic term on the plane $\rho=0$,  $\mathcal{X}_\mathrm{num}$,
    vs
   the same quantity for  FJBD theory  $\mathcal{X}_{\Psi}$, and two approximations described in Sec. \ref{sec:analy}  (with a Gaussian source). In more detail, both the 
 linear superposition of the two one-particle solutions
 $\mathcal{X}_{\mathrm{sup}}$ and the irrotational approximation $\mathcal{X}_{\mathrm{irr}}$ provide an excelent agreement with the numerical results.  Furthermore, the difference between these approximations and the numerical solutions, and hence the importance of the solenoidal component,  is of the order of the numerical error in this regime. The comparison with  FJBD theory demonstrates that the screening is active inside the screening region of the individual objects (shaded region), and that outside  the individual screening regions the scalar gradient is not suppressed. Note also that the kinetic term is significantly suppresed (although not zero in contrast to the isolated object, Sec. \ref{sec:an_sph}), and thus the theory is in the linear regime, near the center of the source. 
 
\begin{figure}
   	 \centering
   	 \includegraphics[width=0.48\textwidth]{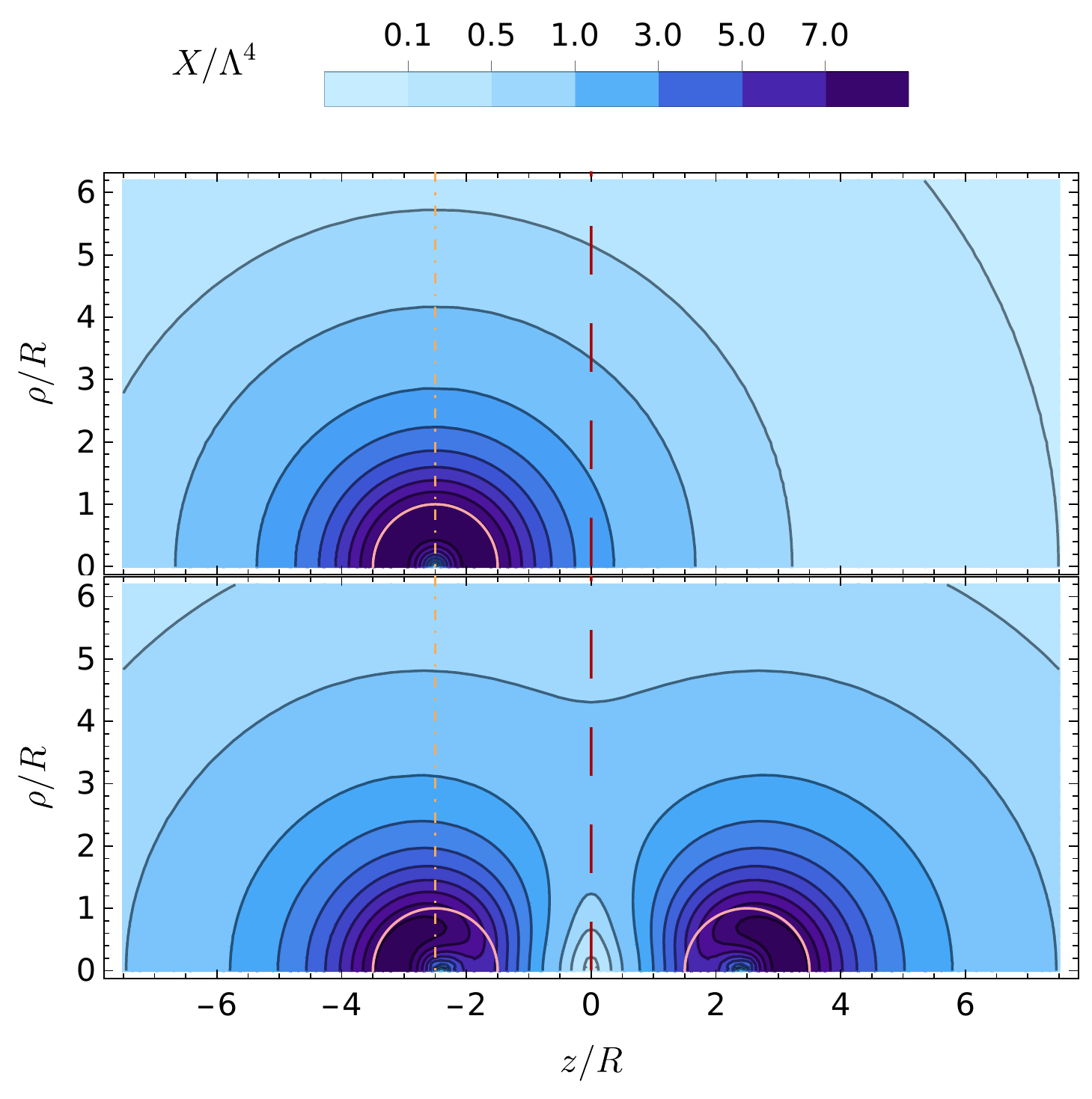}
   	 \caption{Contour plot of the scalar kinetic energy $X$ in the $(\rho,z)$ place.  The bottom panel is for an equal-mass binary with $r_{\rm sc}=9.5 R $ and $D=5R$. The top panel is for just one of the two bodies. The pink semicircles denote the effective radii of the Gaussian source model for the point particles. The orange dot-dashed line connects the geometric center of the left object in the two subplots, while  the red dashed line connects the origins (which are placed at the center of mass of the binary).}
   	 \label{fig:X_rho_z}
    \end{figure}

 More interesting is a scenario where $\kappa > 1$. In Fig.~\ref{fig:2pt_X} (right) we show the same quantities as for the previous case. In order to appriciate the full spatial behaviour of the scalar kinetic energy we also present in Fig.~\ref{fig:X_rho_z}  a contour plot of $\mathcal{X}_{\rm num}$ for one such case (bottom), together with the same plot for the corresponding one-particle case (top). Consider the region around $z=\rho=0$, which is inside the single-particle screening region. In the binary problem, this region corresponds to the saddle point of the scalar profile, where the fifth forces cancel each other  and thus the scalar gradient is suppressed (see the next Sec. \ref{sec:kbub} for further discussion). 
 This can be clearly seen in the bottom panel, where 
 the contour lines
get deformed to allow for near zero gradients in the saddle region.
On the ``outer'' side of the binary, the profile is much closer to the expectation from the single-particle case, although both the scalar gradients and the screening radius become  larger. This is expected, as sufficiently far away from the constituent objects, the system behaves as a composite single object $m_a+m_b$.
Both the irrotational and linear superposition approximations considered earlier capture the essential characteristics of how the screening operates in a binary system as is clear from Fig.~\ref{fig:2pt_X} (right). The superposition ansatz $\mathcal{X}_{\rm{sup}}$ tends to overcorrect the difference between the one-particle case and the two-body by drastically reducing the 
peaks inside the binary 
and enhancing those outside it.  On the other hand, the irrotational approximation $\mathcal{X}_{\rm{irr}}$ makes these adjustments  closer to the true (numerical) solution. 

\begin{figure}
   	 \centering
   	 \includegraphics[width=0.48\textwidth]{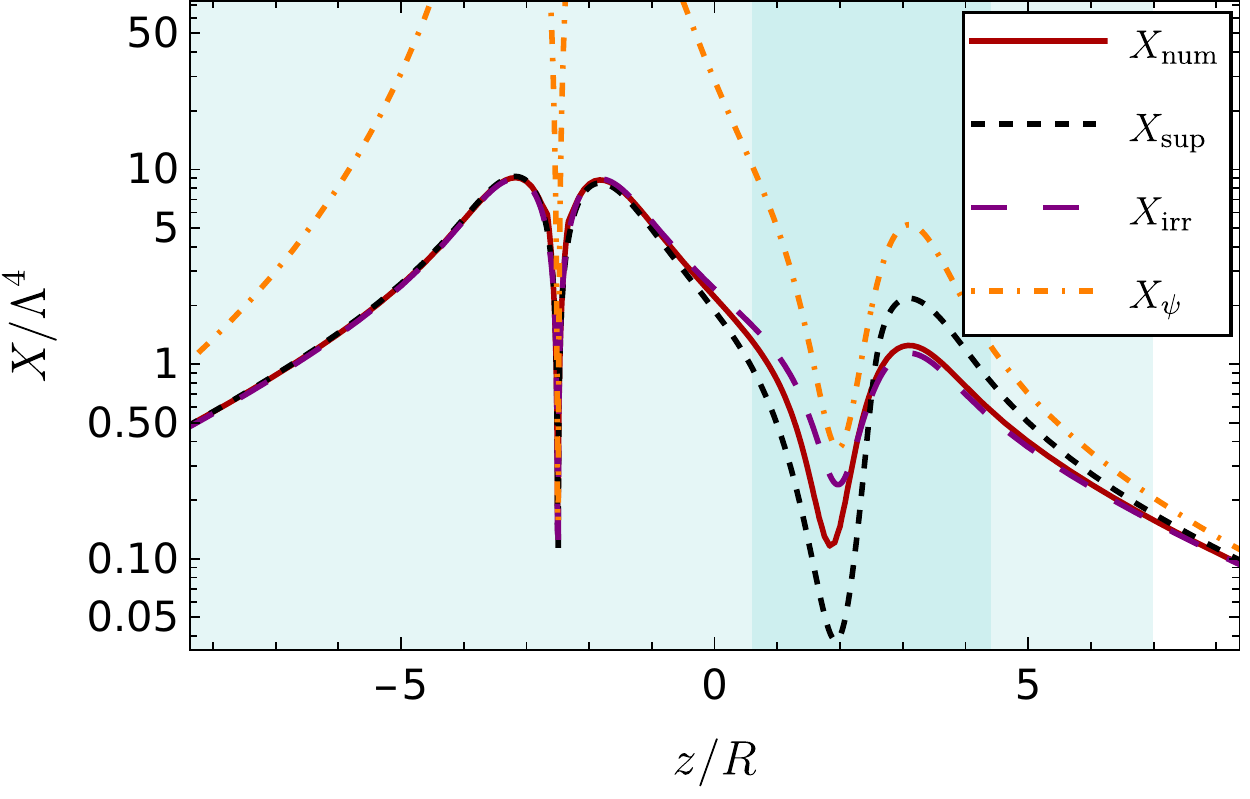}
   	 \caption{The same as in Fig.~\ref{fig:2pt_X} (right), but for $q=25$. The two bodies are placed at $z=\pm 2.5 R$.}
   	 \label{fig:2pt_X_q}
    \end{figure}

Consider now a case with $\kappa > 1 \,, q \gg 1$, shown in Fig.~\ref{fig:2pt_X_q}. As discussed in Sec.~\ref{sec:analy}, both approximations are much closer to the numerical result than in the case of equal-mass systems. In particular, the discrepancy between the analytic approximations and the numerical result is most pronounced in the vicinity of the smaller object. Again, the irrotational approximation is outperforming the simple linear superposition of the one-body solutions.

\begin{figure}
    	\centering
    	\includegraphics[width=0.48\textwidth]{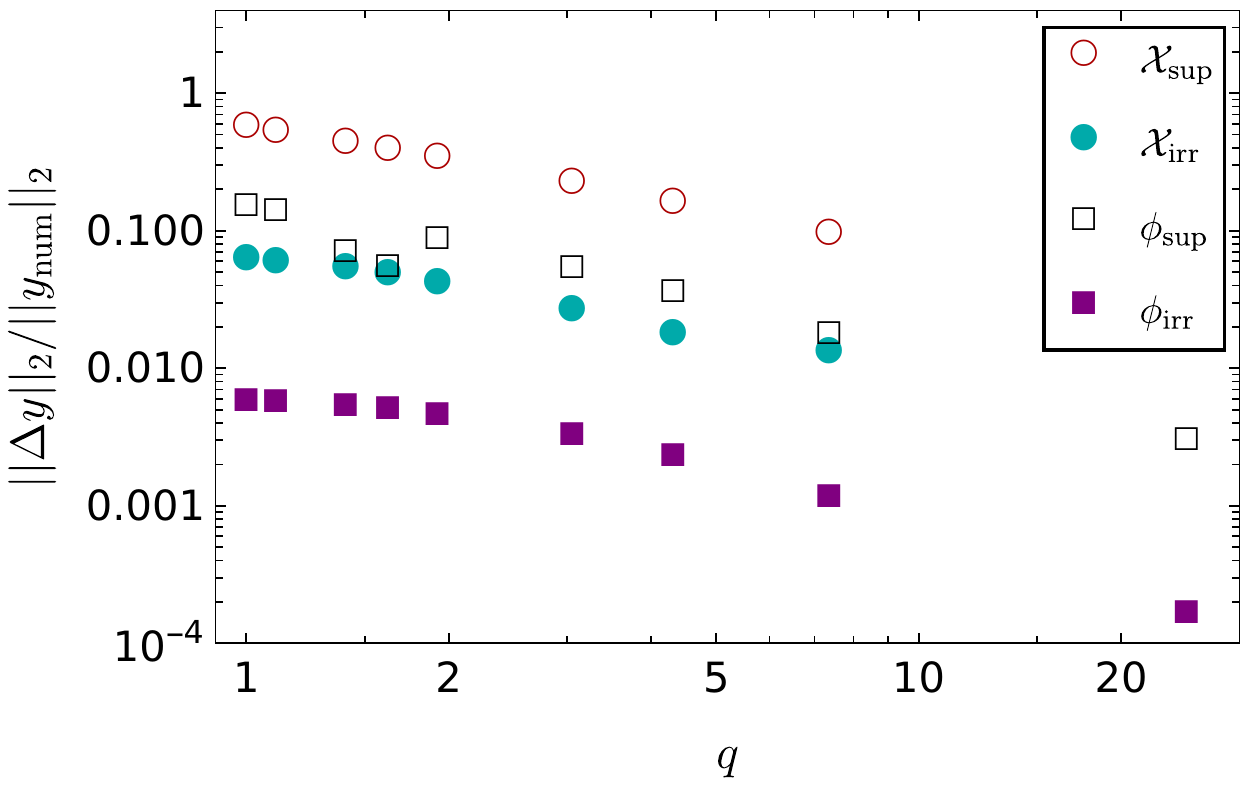}
    	\caption{Relative difference of  the linear superposition and irrotational  approximations from the numerical results, in terms of the $L^2$ norm defined in the text
     and as a function of the mass ratio $q$. The differences are shown 
     for the scalar field $\phi$ and its kinetic energy
     $X$, in the deep screening regime. The system considered is an  $r_{\rm sc}=9.5R$ and $D=5R$.}
    	\label{fig:rel_err_n}
	\end{figure}

In order to compactly describe the two approximations across the mass ratio parameter space, we  define  the following $L^2$ norm
\begin{eqnarray}
||\Delta y||_2 \equiv \sqrt{ \int d\mathsf{V}(y_\mathrm{num}-y_\mathrm{an})^2}
\end{eqnarray}
where $y=\{\phi, \mathcal{X}\}$, with subscripts $\{\mathrm{num},\mathrm{an}\}$, denote the numerical result and the analytic approximation, respectively, and the integral is taken over whole grid. The estimate is sensitive to the non-linear regime, because in the linear regime both approximations give a very good description of the numerical results.  Results are shown in Fig. \ref{fig:rel_err_n} for a  scenario  representative of the deep screening regime $\kappa \gg 1$. We find that the irrotational approximation is outperforming the superposition approximation for all mass ratios, although the relative error of both approximations increases as $q \to 1$. This is completely in line with the conclusions from Sec. \ref{sec:analy}. We also find that the relative error of the kinetic energy saturates at $\sim 10 \%$ for $q=1$. Thus, even in the case of equal masses, the irrotational approximation provides a decent quantitative description of the scalar profile. It is also apparent that, for a given  approximation, the error is smaller for the field than for $X$. A similar phenomenon is observed in the two-body problem for cubic Galileons (where the Vainshtein screening operates~\cite{Nicolis:2008in,Hui:2009kc,Andrews:2013qva,Bloomfield:2014zfa,Joyce:2014kja}), when comparing the superposition approximation and  numerical results~\cite{White:2020xsq}.

    %%%%%%%%%%%%%%%%%%%%
    %%%%%%%%%%%%%%%%%%%
    \subsection{Descreened bubbles} \label{sec:kbub}
    %%%%%%%%%%%%%%%%%%%%%
    %%%%%%%%%%%%%%%%%%%%

In  systems where only attractive forces act, there may be special points where all the forces cancel. For  spherical objects in isolation, one such point is the center, while in $N$-body systems these are the saddle points,  where 
the gradient of the potential vanishes. In theories where the screening is activated by the magnitude of the scalar gradient,
these saddle points (and their neighborhoods) are therefore 
in the perturbative regime, which 
 leads to a possible breakdown of the screening (Fig. \ref{fig:X_rho_z}). 
  This, in turn, suggests that 
  saddle points and their vicinities
may be useful testing grounds for  theories with screening mechanisms. This was recognized for MOND~\cite{Bekenstein:2006fi}, and it was even  suggested that LISA Pathfinder could be used to probe the MOND interpolating function 
if directed towards the saddle point of the solar system~\cite{Magueijo:2011an}.

Following Ref.~\cite{Bekenstein:2006fi},
let us then consider a region where the theory dynamics is in the linear regime near the saddle point. Therefore, we can use the 
superposition of the FJBD scalar gradients (along the axis that connects the two bodies)
to compute  fifth force (per unit mass) as (restoring physical units)
\begin{eqnarray}
(\bm{\nabla} \psi)_z= \frac{\alpha}{4 \pi M_{\rm Pl}} \frac{m}{(z+D/2)^2}- \frac{\alpha }{4 \pi M_{\rm Pl}} \frac{m q^{-1}}{(z-D/2)^2}\,,
\end{eqnarray}
The (saddle) point where the total scalar gradient is zero is given by
\begin{eqnarray}
z_{\mathrm{SP}} = \frac{D}{2} \frac{\sqrt{q}-1}{\sqrt{q}+1} \,.
\end{eqnarray}
Taylor-expanding the scalar gradient around this saddle point, we find the force in its vicinity\footnote{One can further verify that the Hessian matrix is indefinite and thus $(0,z_\mathrm{SP})$ is indeed a saddle point. For a general discussion see Ref.~\cite{Pucci} (we thank Áron Kovács for pointing out this reference).} to be
\begin{eqnarray}
	 \bm{\nabla} \psi_\mathrm{SP} &\approx& A \left((z-z_\mathrm{SP}) \hat{\bm{z}} - \frac{1}{2} \rho \hat{\bm{\rho }} \right) \,,\\
  A &\equiv&  - \frac{ \alpha}{2 \pi M_{\rm Pl}} \frac{m q^{-3/2}}{D^3 }(1+\sqrt{q})^4\,.
\end{eqnarray}
The condition $|\bm{\nabla} \psi_\mathrm{SP} | = \sqrt{X_\psi}|_\mathrm{SP}\leqsim \Lambda^2$ then defines  
the region where the screening may break down. From this condition, one obtains that the size $\delta$ of this region is given by
\begin{equation} \label{eq:k_bub}
\frac{\delta}{D} \simeq \frac{1}{\kappa} \frac{q^{3/2}}{(1+\sqrt{q})^4} \,,
\end{equation}
i.e. this region  shrinks in both the deep screening regime $(\kappa \gg 1)$ and in the extreme mass ratio limit $(q \to \infty)$.

Although both the GR Newtonian force and the scalar fifth force go to zero precisely at the saddle point, in their vicinity 
they are both non-vanishing, with their precise ratio  depending on the value of $\alpha$. Note that constraints on the time-variation of the effective gravitational constant (in the Jordan frame) from  Big Bang nucleosynthesis and  Lunar Laser Ranging experiments require $\alpha \leqsim 0.1$~\cite{Barreira:2015aea}.  
Considering
 three representative binary systems, i.e. Earth and Moon, Sun and Earth, and Sun and Jupiter, taking $\Lambda\approx 2 \times 10^{-3} \mathrm{eV}$ and $\alpha=0.1$, we obtain $\delta \approx 0.2 \mathrm{km}$, $\delta  \approx 1 \mathrm{km}$ and $\delta \approx 2800 \mathrm{km}$,  respectively. As $\delta \propto \alpha^{-2}$, by reducing $\alpha$ the size of 
 the ``descreening'' region 
 grows, but the correction to the GR Newtonian force from the fifth force  decreases by the same amount.

Precise modeling of 
the dynamics near the saddle point of the solar system is challenging,
as it would require accurate ephemeris data~\cite{NANOGrav:2020tig,Fienga:2023ocw}
and even account for the effect of the spacecraft carrying the accelerometer itself.
While this is outside the scope of this work, let us comment on a few parallels with MOND, where these problems have been analyzed to some extent~\cite{Bekenstein:2006fi,Magueijo:2011an}.
First, note that if we had used $\alpha=1$ in our estimates for $\delta$
for the Earth and Moon, Sun and Earth, and Sun and Jupiter systems, they would have differed only by a factor $\sim$ a few from the estimate for MOND in Ref.~\cite{Bekenstein:2006fi}. The reason is that the scale of the MOND critical acceleration is $a_o \approx H_0/6$, thus leading to the same parametric scaling as the cosmologically motivated $k$-essence. In more detail, the MONDian behavior
is triggered by the condition
$a_0 \simeq a_\mathrm{N}$ [where $a_\mathrm{N} \simeq m/(D^2 M^2_{\rm Pl})$],
which is equivalent to the $k$-essence $\partial_r \psi \simeq \Lambda^2$
deep screening condition.

Note that although MOND is not a well-defined theory by itself,  several attempts have been directed at constructing a field theory that can develop a MONDian phenomenology~\cite{Famaey:2011kh}, including hybrid models such as superfluid DM~\cite{Berezhiani:2015bqa}. Implementations that are of $K(X)$ type combine both screening around  matter sources and anti-screening (i.e. enhancement of the scalar gradient, see Sec. \ref{sec:a-sc}) in the low-acceleration regime. Thus, MOND saddle point regions can  be larger than the simple estimate given above, and the fifth force may even dominate the Newtonian force inside them~\cite{Bekenstein:2006fi}. This makes saddle points a potentially better probe of MOND than 
$k$-essence (although not all MOND interpolating functions can be further constrained in this way~\cite{Hees:2015bna}).

%%%%%%%%%%%%%%%%%%%%
	%%%%%%%%%%%%%%%%%%%
	\subsection{Two-body energy and the fifth force} \label{sec:force}
	%%%%%%%%%%%%%%%%%%%%%
	%%%%%%%%%%%%%%%%%%%%
    
In a time-independent system such as the one that we consider, the Hamiltonian (density) is 
given by $\mathcal{H}= -\mathcal{L}$, where the Lagrangian (density) $\mathcal{L}$
is obtained from Eq. (\ref{eq:action}) [specializing to Minkowski space]. From this, one can find the potential energy $E=\int dV \mathcal{H}$ as a function of the system's parameters and the inter-particle separation (see Sec. \ref{sec:analy}).  From the energy, the magnitude of the fifth force between the two particles can then be found as
\begin{equation} \label{eq:5_force_def}
F=\frac{\partial E}{\partial D} \,.
\end{equation}
    
	Let us  consider the  polynomial kinetic function of Eq.~\eqref{eq:function_poly}, which yields
    \begin{eqnarray} \label{eq:ener_poly}
       E =- \int dV \Big[ - \Lambda^4 \sum^N_{n=1} \frac{1}{2n} \left(\frac{X}{\Lambda^4} \right)^n+ \frac{\alpha}{M_{\mathrm{Pl}}}\varphi T \Big]  \,.
    \end{eqnarray}
	Using the equation of motion \eqref{eq:KG_stationary}, we can rewrite this integral as 
    \begin{eqnarray} \label{eq:ener_poly_2}
        \mathcal{E} \equiv \frac{E}{D^3\Lambda^4}= -\int d\mathsf{V} \sum^N_{n=1} \Big(\frac{2n-1}{2n} \Big) \mathcal{X}^n\,,
    \end{eqnarray}
where we have also used the rescaling of Eq.~\eqref{eq:tildas_units}.
	Noting that in the deep screening regime the highest power of $\mathcal{X}$  dominates the integral and using
 the irrotational approximation,
 one can apply Eq. \eqref{eq:X_X_psi}
 and obtain that the energy $\mathcal{E}_{\mathrm{sc}}$ of the screened regions ($\mathrm{sc}$) is
 given by
    \begin{eqnarray} \label{eq:ener_sc}
 \mathcal{E}_{\mathrm{sc}}
        \approx -\kappa^{\frac{2N}{2N-1}}  \Big(\frac{2N-1}{2N} \Big)\int_\mathrm{sc} d\mathsf{V}  \hat{\mathcal{X}}_\Psi^{N/(2N-1)} \,.
    \end{eqnarray}
 The total  energy is then obtained by adding the subdominant term $\approx -\int_{\mathrm{un-sc}} d\mathsf{V} \mathcal{X}_\Psi/2$ that corresponds to the
 `unscreened' region 
$\mathrm{un-sc}$.
[Note indeed that the integral in Eq.~\eqref{eq:ener_sc}  diverges in the unscreened region].    One can observe  that for $N=1$, we recover the FJBD scaling $ E  \propto D^{-1}$, while for $N=2$ we obtain $E \propto D^{1/3}$, as expected on dimensional grounds and from the single particle limit. Note also that  in contrast to the Newtonian/FJBD case, the scalar self-energy does not diverge in the point-particle limit thanks to the screening.   From the scaling of the source of the solenoidal component [Eq.~\eqref{eq:F_k_deep_scr}], we find that $\mathcal{B} \propto \kappa$ when $\kappa \gg 1$. Thus, including the solenoidal component  in Eq.~\eqref{eq:ener_poly_2} does not change the overall scaling of the energy  with $\kappa$ in the deep screening regime, as given by Eq.~\eqref{eq:ener_sc}.
    
 The amplitude of the fifth force [from Eqs. \eqref{eq:5_force_def} and \eqref{eq:ener_sc}] is then given by
 \begin{eqnarray} \label{eq:5_f_sc}
    \mathcal{F}_\mathrm{sc} &\equiv &\frac{F_\mathrm{sc}}{D^2\Lambda^4}  \approx -\kappa^{\frac{2N}{2N-1}} I_N(q) \,, \nonumber \\
  I_N(q) &\equiv& \frac{1}{2}\int_\mathrm{sc} d\mathsf{V}   \hat{\mathcal{X}}_\Psi^{(1-N)/(2N-1)} \partial_\mathsf{D}  \mathcal{\hat{X}}_\Psi\big|_{\mathsf{D}=1} \,.
 \end{eqnarray}
 This indicates a clear suppression when $N>1$ in comparison  to the FJBD limit $N=1$. For instance,  for $N=1$
 one has $F_\mathrm{sc}  \propto D^{-2}$ (Newton's law), while for $N=2$ one obtains $F_\mathrm{sc}  \propto D^{-2/3}$. Unlike the energy, the fifth force diverges in the limit $D \to 0$. However,
 this is simply an artifact of the point-particle model, i.e. it disappears for extended sources (see e.g. Ref.~\cite{terHaar:2020xxb}
 and the discussion 
in our Sec. \ref{sec:an_sph}, App. \ref{app:coloumb}). The details of the calculation of $I_N(q)$ are presented in App. \ref{app:5force_calc}.

\begin{figure}
    	\centering
    	\includegraphics[width=0.48\textwidth]{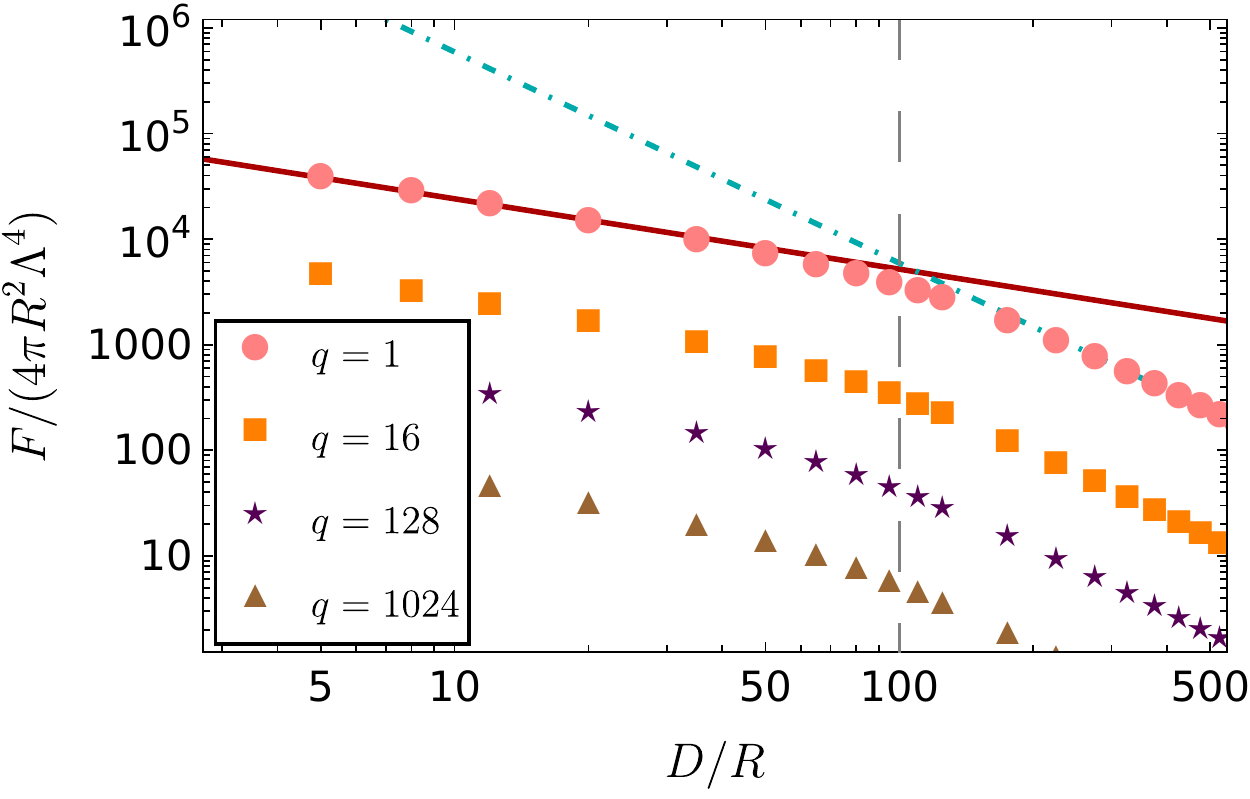}
    	\caption{Fifth force $F$ for four values of the mass ratio $q$, calculated using the irrotational approximation  and various values of $D/R$, for fixed $r_{\mathrm{sc}}/R=100$ (denoted by the gray dashed line). For $q=1$,
     we also show the FJBD limit [Eq. \eqref{eq:force_psi}, cyan dot-dashed line] as well as the deep screening approximation [Eq. \eqref{eq:5_f_sc}, solid red line]. }
    	\label{fig:5force}
	\end{figure}

In order to verify the internal consistency of
our irrotational approximation (in both the 
 deep screening and  FJBD regimes),   we have calculated the fifth force [from Eqs. \eqref{eq:5_force_def} and \eqref{eq:ener_poly_2}] semi-analytically in quadratic $k$-essence (see App. \ref{app:5force_calc} for  details). In Fig. \ref{fig:5force}, we show how the magnitude of the fifth force  depends on the mass ratio and on the inter-particle separation in units of the object's radius (although we stress that our results do not depend on the details of the object's internal structure, as long as one focuses on the object's exterior). For $q=1$, we also show, for comparison,  
 the FJBD limit  $F_\psi$ and the deep screening limit given by Eq.~\eqref{eq:5_f_sc} (which suppresses the fifth force relative to FJBD theory). 
  As can be seen, the change between the two regimes is abrupt, for all mass ratios, and  the two approximations provide a very good description of the scaling of the fifth force with distance  
  in a piecewise fashion.

\begin{figure}
    	\centering
    	\includegraphics[width=0.48\textwidth]{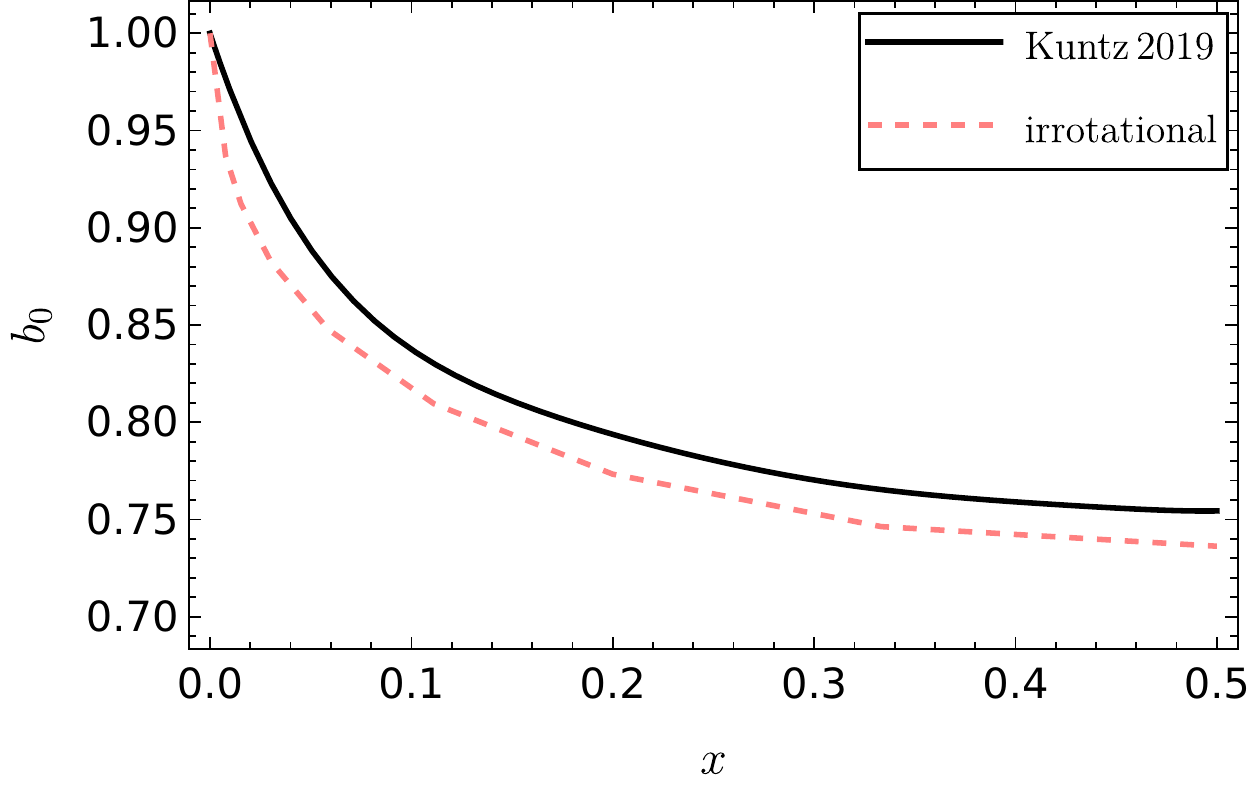}
    	\caption{Ratio of the fifth force amplitudes  between the  two point particles and in the test-mass limit, in quadratic $k$-essence, calculated from the numerical simulation of Ref.~\cite{Kuntz:2019plo} (black) and using the irrotational approximation in the deep screening regime (pink, dashed).}
    	\label{fig:b0x}
	\end{figure}

Having established that the deep screening limit  of Eq.~\eqref{eq:5_f_sc} is valid in the context of the irrotational approximation, we can compare that limit with the full numerical result of Ref.~\cite{Kuntz:2019plo}  for quadratic $k$-essence.  Let us define the force in the test-mass limit by performing the standard Newtonian reformulation of a two body problem  into the motion of a fictitious particle with the reduced mass $\mu=m_a m_b /(m_a+m_b)$ around a particle with the total mass $m_a+m_b$ (see e.g. Ref.~\cite{poisson_will_2014}). The amplitude of the force is then given by $F_\mathrm{tm}=(\alpha/M_{\rm Pl}) \mu \, \partial_r \varphi|_{r=a}$. Thus, from Eq. \eqref{eq:KG_poly_N_sc} [and using the rescaling of Eq.~\eqref{eq:tildas_units}] we  obtain
 \begin{eqnarray}
\frac{\mathcal{F}_\mathrm{tm}}{4\pi} =  \left[\frac{\kappa^4}{ q (q+1)^2} \right]^{1/3}  \,.
 \end{eqnarray}
From the irrotational approximation, it follows that the force in the deep screening regime [Eq.~\eqref{eq:5_f_sc}] has the same scaling with $\kappa$ as the test-mass limit (and the solenoidal component does not change this scaling, as argued above).
Therefore, the ratio of $F_{\rm sc}$ and $F_{\rm tm}$ depends only on the mass ratio $q$. The same conclusion was reached in Ref.~\cite{Kuntz:2019plo} using an effective-one-body approach.
 Following Ref.~\cite{Kuntz:2019plo}, let us define
 \begin{eqnarray}
  	F_\mathrm{sc} = b_0(x) F_\mathrm{tm} \,, \quad
  	x=\frac{1}{1+q}\,,
 \end{eqnarray}
where we expect that  $\lim_{x \to 0} b_0 \to 1$.  We have found  $b_0(x)$ [from $I_2(q)$] semi-analytically, 	and we have compared it with the fit of the full numerical result from  Ref.~\cite{Kuntz:2019plo} (Eq. 46) in Fig.~\ref{fig:b0x}. As can be seen, this comparison confirms the observation from  Ref.~\cite{Kuntz:2019plo} that the screening is more efficient in  equal-mass systems 
than in the extreme-mass ratio limit. (This implies a breakdown of the weak equivalence principle, as further elaborated in  Ref.~\cite{Kuntz:2019plo}). Note that the relative enhancement of the screening  
relative to the extreme mass-ratio limit is at most $\sim 25\%$. Also note that our findinds confirm that the irrotational approximation is in good agreement with the full numerical results,  differing only by a few percent from the latter.

Finally, let us emphasize that the breakdown of screening in the descreened bubbles (see Sec. \ref{sec:kbub}) is a \textit{local} phenomenon, which can be probed with a third test body. Descreened bubbles can in principle affect the two-body energy and the fifth force, since the latter are expressed as integrals over  all space. For a  given $\kappa$, 
Eq. \eqref{eq:k_bub} predicts that the largest descreened bubbles appear 
for $q \approx 9$. However, as  clear from Fig. \ref{fig:b0x}, in the deep screening regime this does not 
significantly impair
the efficiency of the screening 
 mechanism.
 
    %%%%%%%%%%%%%%%%%%%%
    %%%%%%%%%%%%%%%%%%%
    \section{Other theories} \label{sec:k-other}
    %%%%%%%%%%%%%%%%%%%%%
    %%%%%%%%%%%%%%%%%%%%

 Thus far, we have focused on the polynomial form of the $k$-essence kinetic functon, given by Eq.~\eqref{eq:function_poly}. Let us now broaden the scope of the possible kinetic functions, 
 and investigate again the 
 non-relativistic limit for single bodies and for binary systems. We will employ the same notation as in the previous sections.
Note that while in polynomial $k$-essence  a generic astrophysical system is in the regime of validity of the EFT when the screening operates, 
this may not be the case for some theories considered in this section, and in particular opposite DBI and anti-screening theories (see App. \ref{app:eft}). 

%%%%%%%%%%%%%%%%%%%%
 \subsection{Beyond (simple) polynomial $k$-essence} \label{sec:kin_f_var}  
%%%%%%%%%%%%%%%%%%%%%%

One may wonder how   values of the polynomial coefficients different from those in Eq.~\eqref{eq:function_poly} impact our previous discussion and results. 
Since
the highest power of $X$
dominates upon the others in the deep screening regime,
our results should be qualitatively unchanged (as long as the coefficient of the highest power of $X$ has the sign leading to screening in the first place).
 We have performed simple numerical experiments to check this, e.g. we have considered a sextic polynomial $ \mathcal{K}_6=-\mathcal{X}/2+ c_2 \mathcal{X}^2 + c_4 \mathcal{X}^4 -\mathcal{X}^6/12$, with randomly generated  values of the coefficients $\{c_2,c_4\}$ in the interval $-5 \leq c_i \leq 5$ but requiring that the condition of Eq.~\eqref{eq:X_cond_invert} is satisfied. We have calculated the suppression factor $\mathcal{X}/\mathcal{X}_\psi$ for the isolated Gaussian source as in Sec. \ref{sec:an_sph} for several such realizations. Comparing with our default model $ \mathcal{K}_6=-\mathcal{X}/2-\mathcal{X}^6/12$, we have found that outside the object but within the screening region, the suppression factor  varies by at most a few percent. The differences between the various realizations peak around the screening radius, where they reach $\sim 10 \%$, as  this is the transition region where the effect of the subleading terms $\mathcal{X}^2, \mathcal{X}^4$ is maximized. The relative difference is further suppressed outside the screening region, when the FJBD limit is asymptotically approached. We expect this conclusion to hold also for the two-body problem.

If one relaxes the assumption of a polynomial kinetic function, one can also engineer particular functions  passing different cosmological and solar system constraints while still providing a viable scalar-tensor theory of gravity~\cite{Brax:2014gra,Barreira:2015aea}. One such model was considered in Ref.~\cite{Barreira:2015aea}
and is given by
\begin{eqnarray} \label{eq:K_arctan}
    \mathcal{K}_{\tan^{-1}} = - 1 - \frac{\mathcal{X}}{2}  - \mathcal{K}_\star \left[\mathcal{X} - \mathcal{X}_\star \arctan{\left(\frac{\mathcal{X}}{\mathcal{X}_\star} \right) } \right] \,,
\end{eqnarray}
where $\{\mathcal{X}_\star,\mathcal{K}_\star\}$ are free parameters of the model. Note that $\mathcal{K}'(\mathcal{X}) \to \mathcal{K}_\star$ as $\mathcal{X} \to \infty$ (see Ref.~\cite{Brax:2016jjt} for the quantum aspects of this model). In the irrotational approximation, one cannot invert Eq.~\eqref{eq:X_cond_invert} exactly, although in the deep screening regime the relation is approximately linear,  $\mathcal{X} \approx \mathcal{X}_\Psi /(1 + \mathcal{K}_\star)^2$, and the suppression of the fifth force is realized through a large value for $\mathcal{K}_\star$. We show a numerical solution $\mathcal{X}(\mathcal{X}_\Psi)$ for $\mathcal{X}_\star = -2$, $\mathcal{K}_\star = 10^3$ (model I from Ref.~\cite{Barreira:2015aea}) in Fig. \ref{fig:X_Xpsi}.

In the context of  the two-body problem, let us assess the importance of the solenoidal component using the arguments of Sec.~\ref{sec:analy}. The quantity that characterizes the non-linearities, i.e. $\mathcal{N}_K$ [Eq. \eqref{eq:Fdel}], is given in the deep screening regime by
\begin{eqnarray} \label{eq:FK_atan}
\mathcal{N}_K \approx \kappa^{-3} \mathcal{K}_\star (1+\mathcal{K}_\star)^3 \mathcal{X}_\star^2 \frac{|\bm{\nabla} \hat{\mathcal{X}}_\Psi|}{\hat{\mathcal{X}}_\Psi^{5/2}} \,.
\end{eqnarray}
As the screening  arises in this model from a large factor $\mathcal{K}_\star$, the relative strength of the solenoidal and  irrotational components is controlled by $(\mathcal{K}_\star/\kappa)^4$ [we remind the reader that $S_\Psi \propto \kappa$, see Eq. \eqref{eq:S_psi}]. This scaling  makes the solenoidal component much more suppressed than in the case of a polynomial kinetic function, for the parameter values considered in Ref.~\cite{Barreira:2015aea}. 

 \subsection{``Opposite" DBI} \label{sec:dbi} 

Let us consider a class of models where the scalar gradient $\mathcal{X}$ saturates in the strongly interacting regime. A particularly interesting model is 
the (opposite) DBI one~\cite{Dvali:2010jz,Burrage:2014uwa,deRham:2014wfa}
	\begin{equation} \label{eq:k_dbi}
    	\mathcal{K}_{\mathrm{DBI}}= \sqrt{1-\mathcal{X}/2} \,.
	\end{equation}
Standard DBI theory (obtained by flipping the overall sign of the Lagrangian and the sign in front of $\mathcal{X}$) can be embedded in string theory,
but does not allow for screening~\cite{Brax:2012jr,Brax:2014gra}. However, the ``opposite DBI'' kinetic function in Eq.~\eqref{eq:k_dbi} may appear naturally in higher dimensions and possesses a higher symmetry group than the standard shift-symmetric $K(X)$ theory analyzed thus far~\cite{deRham:2014wfa,Burrage:2014uwa}.

\begin{figure}
  	  \centering
  	  \includegraphics[width=0.48\textwidth]{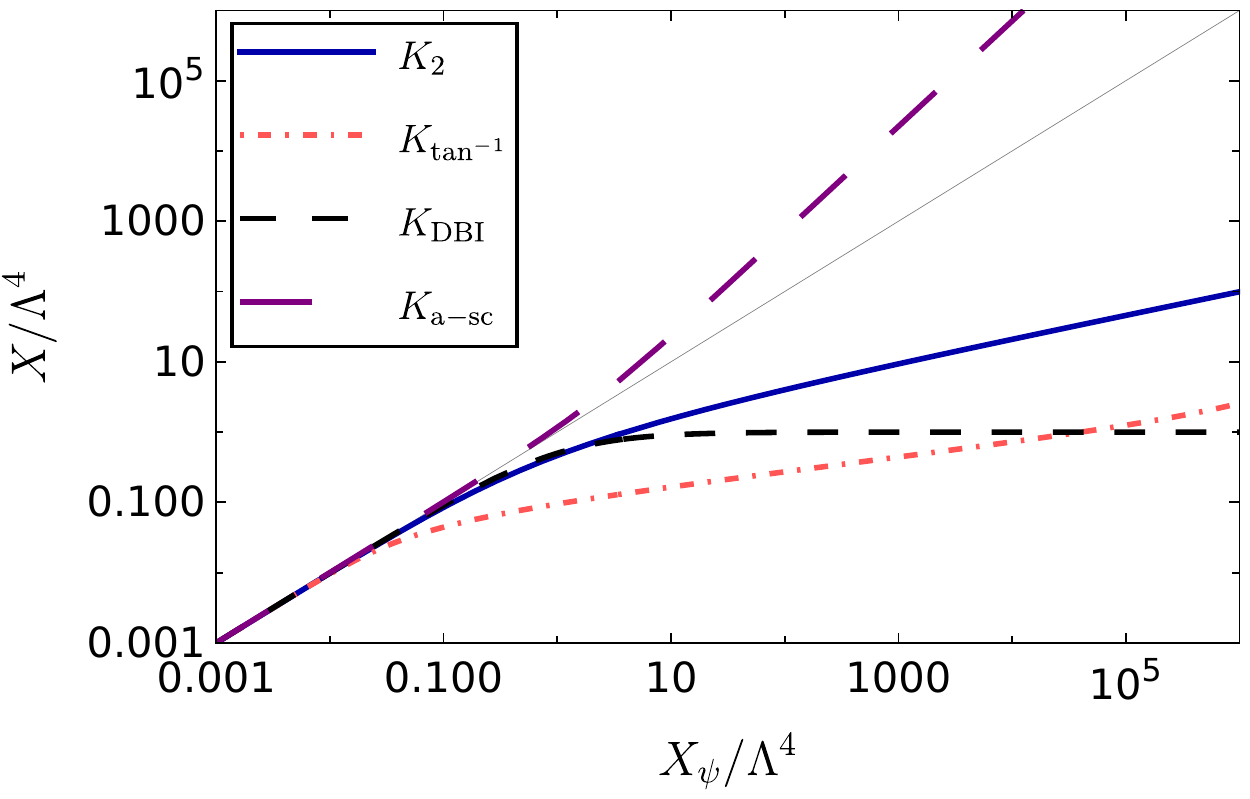}
  	  \caption{Relation between
     the kinetic energy $X$ and the FJBD kinetic energy
     $X_\psi$ in the irrotational approximation, for several choices of the kinetic function. The latter include
    quadratic $k$-essence $K_2$ [Eq. \eqref{eq:K_2}]; the $\rm{arctan}$ model $K_{\tan^{-1}}$   [Eq. \eqref{eq:K_arctan}; $\mathcal{X}_\star=-2$, $\mathcal{K}_\star=10^3$]; opposite DBI theory  $K_{\mathrm{DBI}}$  [Eq. \eqref{eq:k_dbi}]; and the anti-screening model $K_\mathrm{a-sc}$ [Eq. \eqref{eq:k_anti}; $p=5/6$]. The thin solid line corresponds to $X=X_\psi$.}
  	  \label{fig:X_Xpsi}
\end{figure}

\begin{figure}
  	  \centering
  	  \includegraphics[width=0.48\textwidth]{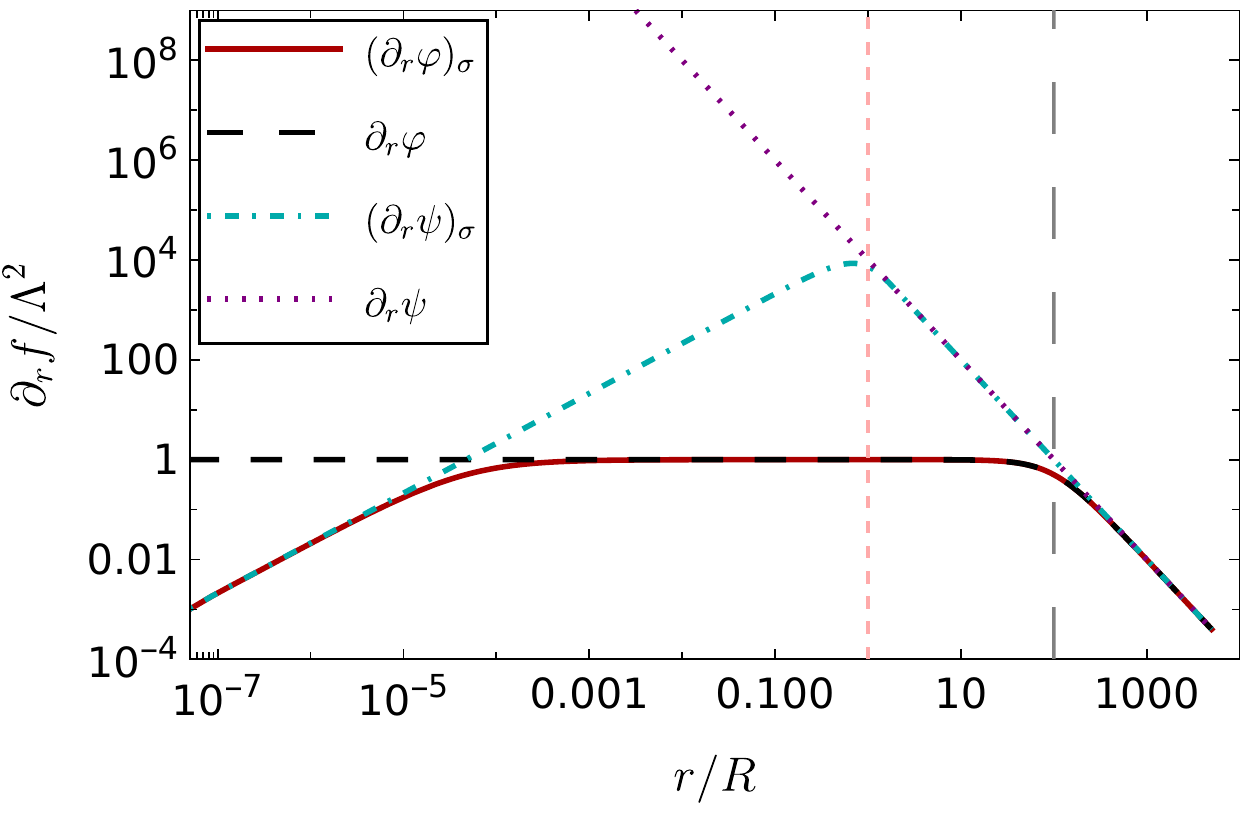}
  	  \caption{The same as in Fig.~\ref{fig:1pt_X}, but for (opposite) DBI theory.
  	  \label{fig:1pt_DBI}}
\end{figure}

Like in the case of quadratic $k$-essence, the solution for a single point-particle source can be expressed in terms of the hypergeometric function:
  \begin{eqnarray} \label{eq:1P_DBI}
        \phi &=& 1.85 \mathsf{r}_{\rm{sc}}  - \mathsf{r} \,\prescript{}{2}{F}_1 \Big[\frac{1}{4},\frac{1}{2};\frac{5}{4};-\Big(\frac{\mathsf{r}}{\mathsf{r}_{\rm sc}}\Big)^4\Big]\,. \label{eq:1p_sol_dbi}
    \end{eqnarray}
This exact solution features a screening radius $ \mathsf{r}_{\rm{sc}}= \sqrt{\kappa}$. Like in Sec. \ref{sec:an_sph}, we show the scalar gradient for an isolated Gaussian source and for a point particle in Fig. \ref{fig:1pt_DBI}. In contrast with the polynomial $k$-essence, now even in the point-particle case the scalar gradient does not diverge at the particle's location, and 
the point-particle and Gaussian models are much closer even inside the source, down to very small radii.  We also find that the screening in opposite DBI is more efficient than in quadratic $k$-essence (see Figs.~\ref{fig:1pt_X} and~\ref{fig:1pt_DBI}, and also Fig. \ref{fig:X_Xpsi}).

Away from spherical symmetry (or other highly symmetric configurations), the solenoidal component is in general not zero in opposite DBI, unlike what was implicitly assumed in Ref.~\cite{Burrage:2014uwa}. However, we can consider the irrotational approximation described in Sec. \ref{sec:analy} (see Fig. \ref{fig:X_Xpsi}) and obtain
	\begin{equation}
\mathcal{X}=\frac{\mathcal{X}_\Psi}{1+\mathcal{X}_\Psi} \,.
	\end{equation}
We can then calculate the source of the solenoidal field that encodes non-linearities in the two-body problem [see Eq. \eqref{eq:Fdel}], obtaining
 \begin{eqnarray}
 \mathcal{N}_K \approx \frac{1}{4} \sqrt{\mathcal{X}(1-\mathcal{X})} \,.
 \end{eqnarray}
One can see that as $X$ flattens  in the deep screening regime, $\mathcal{N}_K \to 0$. Thus, the solenoidal component in opposite DBI is even more suppressed than for a polynomial kinetic function.

 \subsection{Anti-screening} \label{sec:a-sc}

As noted in Sec. \ref{sec:intro}, a drawback of  generic $k$-essence models is the absence of a standard UV completion~\cite{Adams:2006sv,Aoki:2021ffc}. Thus, in Ref.~\cite{Hertzberg:2022bsb} a class of shift-symmetric theories that do not violate positivity bounds\footnote{These models were unfortunately advertized in Ref.~\cite{Hertzberg:2022bsb} as causal modifications of gravity, as opposed to superluminal models that allow for screening. However, as elaborated in Sec. \ref{sec:intro}, causality is not an issue in   superluminal $k$-essence theories~\cite{Armendariz-Picon:2005oog,Bruneton:2006gf,Babichev:2007dw,Brax:2014gra,Bezares:2020wkn,Lara:2021piy}.} and which are thus expected to admit such a completion were considered. Those are described by
\begin{eqnarray} \label{eq:k_anti}
 \mathcal{K}_\mathrm{a-sc} =- \frac{1}{p} \left[\left(1+\mathcal{X} /2 \right)^p -1\right] \,,  
\end{eqnarray}
where $1/2\leq p < 1$ ($p=1/2$ corresponding to the standard DBI theory, see Ref.~\cite{Brax:2014gra}, while $p=1$ yields FJBD theory). These theories, however, lead to \textit{anti-screening} i.e. an enhancement of the fifth force near  matter sources (see Fig. \ref{fig:X_Xpsi}). These theories, like FJBD theory, are then only relevant  if the coupling of the scalar to matter satisfies the Cassini bound. The anti-screening phenomenon could then provide additional constraints in the solar system and in the strong gravity regime~\cite{Hertzberg:2022bsb}.

The Helmholtz decomposition and the arguments of Sec. \ref{sec:EoM_R3}, \ref{sec:analy} can also be applied to theories with anti-screening. In particular, in spherical symmetry the solenoidal component is zero, and $\mathcal{X}(\mathcal{X}_\Psi)$ can be found from Eq. \eqref{eq:X_poly_eq}  (see Fig. \ref{fig:X_Xpsi}).
% (semi-)analytically, e.g. for $p=5/6$ in closed form 
%
Away from spherical symmetry, from Eq. \eqref{eq:Fdel} one finds (e.g. for the special case $p=5/6$)
\begin{eqnarray}
\mathcal{N}_K  \approx  \frac{\kappa}{8} \frac{|\bm{\nabla} \hat{\mathcal{X}}_\psi|}{\sqrt{\hat{\mathcal{X}}_\Psi}}\,,
\end{eqnarray}
i.e. the same as in Eq. \eqref{eq:F_k_deep_scr} for $N=5/6$. Therefore, the details of the anti-screening model are captured in a dimensionless prefactor, and the dependence on $\kappa$ and $q$ is the same as in polynomial $k$-essence (see Fig. \ref{fig:source_q}).

%
 
	%%%%%%%%%%%%%%%%%%%%
	%%%%%%%%%%%%%%%%%%%
	\section{Conclusions} \label{sec:fin}
	%%%%%%%%%%%%%%%%%%%%%
	%%%%%%%%%%%%%%%%%%%%
We have shown that shift-symmetric scalar-tensor theories,  involving only first derivatives in scalars, can be reformulated by covariantly splitting the scalar gradient into a longitudinal $\partial_\mu \psi$ and a transverse component $B_\mu$ (Hodge-Helmholtz decomposition; Sec. \ref{sec:cov}). The longitudinal component reduces to a (free) Klein-Gordon field, while the transverse component obeys a hyperbolic equation with a non-linear source. We have shown that for spherical and static sources, the transverse component identically vanishes (Sec. \ref{sec:EoM_R3}, \ref{sec:an_sph}). In this situation,  the problem  reduces
to solving a linear elliptic equation and then an algebraic one.

 In general, and also in the case of two-body 
 non-relativistic systems that we  consider in this paper,  the solenoidal component does not vanish (unlike what was implicitly assumed e.g. in Refs.~\cite{Gabadadze:2012sm, Burrage:2014uwa}). Outside the screening region of a two-body system (which is controlled by the ratio between the screening radius of the more massive objects $r_{\mathrm{sc}} $ and the inter-particle separation $D$, as well as by the mass ratio $q$), the solenoidal component is perturbativly suppressed and the superposition approximation provides a good description of the full results (see Sec. \ref{sec:analy} and \ref{sec:numerics}). Inside the screening region, we have developed an approximate `irrotational' scheme that starts by ignoring the solenoidal component and  finds the scalar gradients $X$ by solving an algebraic problem. This approach was validated by checking self-consistency (Sec. \ref{sec:analy}), by solving the full system numerically (Sec. \ref{sec:numerics}) and by comparing with the previous results of Ref.~\cite{Kuntz:2019plo}  (Sec. \ref{sec:force}). We have shown in Sec. \ref{sec:analy} that irrespective of the form of the kinetic function, the irrotational field  dominates upon the solenoidal one when $D \gg r_{\mathrm{sc}} $ and/or $q \gg 1$. In these regimes, ignoring the solenoidal component  will introduce only  small errors in the description of the scalar gradients. Furthermore, we have shown that even when $r_{\mathrm{sc}} \gg D$ and $q \approx 1$, the irrotational approximation will introduce only  percent-level errors in the binary fifth force, in comparison with the full numerical results for quadratic $k$-essence (Sec. \ref{sec:force}). Kinetic functions whose growth is suppressed in the deep screening regime will have even more suppressed solenoidal components (Sec. \ref{sec:kin_f_var}, \ref{sec:dbi}). The irrotational approximation can also be applied  for theories that exhibit anti-screening (Sec. \ref{sec:a-sc}).

At the physical level, our results, both analytic and numerical, show that the absence of spherical symmetry in a binary system does not make screening necessarily inefficient.  On the one hand, the non-linear nature of $k$-essence generally renders screening slightly {\it more efficient}, relative to the test-body limit, in  equal-mass systems (Sec.~\ref{sec:force}). This has already been established numerically, and its consequences on violations of the weak equivalence principle elaborated, in Ref.~\cite{Kuntz:2019plo}. However, we show that binaries  also produce ``descreened'' regions near 
the system's saddle point
(Sec.~\ref{sec:kbub}). These regions may in principle be probed in the solar system with sufficiently precise accelerometers. 
A natural continuation of this work would be to assess the validity of the irrotational approximation in $N$-body systems. In relation to $k$-essence probes in the solar system (e.g. with these descreened regions), such an irrotational approach could alleviate the numerical difficulty and cost of full multi-body simulations (for a related numerical study of the Sun-Earth-Moon system in the cubic Galileon theory, see Ref.~\cite{White:2020xsq}).

As the Hodge-Helmholtz decomposition can be implemented in a covariant way, it is an interesting question to consider whether it can be helpful in dynamical problems. For instance, stellar collapse in $k$-essence leads to  a breakdown of screening~\cite{Bezares:2021yek}.  Indeed, due to black-hole no-hair theorems~\cite{Hui:2012qt,Sotiriou:2013qea,Graham:2014mda,Creminelli:2020lxn,Capuano:2023yyh}, the star's scalar ``charge'' must be radiated away~\cite{Bezares:2021yek}, 
producing a  potentially
observable gravitational-wave signal. Another interesting problem is a binary neutron star's inspiral. In scalar-tensor theories, the orbital energy of a binary decreases because of the emission of both scalar and tensor gravitons. In FJBD, and related perturbative theories where the screening does not operate, the binary inspiral can be systematically studied within the PN formalism~\cite{Will:1989sk,Damour:1992we,Bernard:2018hta,Kuntz:2019zef,Bernard:2022noq,poisson_will_2014}. However, in theories with screening, such an approach is not straightforward. Thus far, in theories with strong non-linearities the inspiral has been studied by perturbing the scalar field around the background field generated by a fictitious isolated body located at the center of mass of the system~\cite{deRham:2012fw,Dar:2018dra,Hertzberg:2022bsb}. On the other hand, numerical simulations have been performed in cubic $k$-essence (including GR), scanning the $r_\mathrm{sc}/D \approx 1-6$ range~\cite{Bezares:2021dma}. Both of these problems may benefit from  (a more systematic) analytic approach based on the Hodge-Helmholtz decomposition.

    %%%%%%%%%%%%%%%%%%%%%%%%%%%%%%%%%%%%%%%%%%%%%%%%%%%%%%%%%%%%%%%%%%%%%%%%%%%%%%
    \begin{acknowledgments}
   	 We would like to  thank Miguel Bezares, Marco Crisostomi, Mario Herrero-Valea and Adrien Kuntz  for insightful conversations on the topics of this paper. We acknowledge financial support provided under the European Union's H2020 ERC Consolidator Grant ``GRavity from Astrophysical to Microscopic Scales'' grant agreement no. GRAMS-815673.

    \end{acknowledgments}
    %%%%%%%%%%%%%%%%%%%%%%%%%%%%%%%%%%%%%%%%%%%%%%%%%%%%%%%%%%%%%%%%%%%%%%%%%%%%%%
    
    \appendix

   		 %%%%%%%%%%%%%%%%%%%%%%%%%%%%%%%%%%%%%%%%%%%%%%%%%%%%%%%
	\section{Regime of validity of the Effective Field Theory} \label{app:eft}
	%%%%%%%%%%%%%%%%%%%%%%%%%%%%%%%%%%%%%%%%%%%%%%
    
Naively, it might seem  that the screening regime lies outside the regime of validity of the EFT, as $X \geqsim \Lambda^4$ (see Figs.~\ref{fig:1pt_X} and~\ref{fig:2pt_X}). The right question, however, is  whether quantum effects can induce significant corrections to the classical description of  screening. This can be assessed with the background field method~\cite{Schwartz:2014sze}, by comparing the classical action, evaluated on the screening-inducing field, and the one-loop effective action on top of the same background field~\cite{deRham:2014wfa}.  In spherical symmetry, non-linearities become more important as the radial coordinate decreases (see Sec. \ref{sec:an_sph}). Thus, the EFT description is valid as long as $r \gg r_\mathrm{UV}$, where $r_\mathrm{UV}$ is the scale where  quantum corrections become significant and consequently the UV physics must play a role. These scales are given by
\begin{eqnarray} \label{eq:r_UV}
r^{\mathrm{poly}}_\mathrm{UV} \sim \frac{1}{\Lambda}  \left(\Lambda r_{\mathrm{sc}}\right)^{-N/(N-1)} \,, \\  r^{\mathrm{DBI}}_\mathrm{UV} \sim \frac{1}{\Lambda}  \left(\Lambda r_{\mathrm{sc}}\right)^{2/3} \,,
\end{eqnarray}
for the point-particle screening  in  polynomial $k$-essence [Eq.~\eqref{eq:function_poly}] and DBI [Eq.~\eqref{eq:k_dbi}], in respective order~\cite{deRham:2014wfa}.
Note however that the scalar gradients do not grow  arbitrarily, because at some point the radius of the source (e.g. a star) is reached. As we have seen in Sec. \ref{sec:an_sph}, the maximal value of $X$ is reached at the surface of the object, and $X$ then decreases as one progresses towards the center, ultimately entering the linear regime. Thus, a sufficient criterion to assess whether one is in the EFT regime 
consists of checking whether $r_\mathrm{UV} \approx R$, where $R$ is the effective radius of the source. Let us focus for concreteness  on  quadratic $k$-essence.  From Eqs.~\eqref{eq:1P_sol} and~\eqref{eq:r_UV} one has
\begin{eqnarray} \label{eq:r_UV_2}
	r^{(N=2)}_\mathrm{UV} \approx   10^{-43} \mathrm{km} \left( \frac{\alpha}{0.1}\right)^{-1}   \left( \frac{\Lambda}{\rm{meV}} \right)^{-1}   \left( \frac{m}{M_\odot}\right)^{-1} \,.
\end{eqnarray}
For cosmologically motivated values of $\Lambda$ and any astrophysical object, one is clearly in the regime of validity of the EFT. As an extreme example, let us consider the LISA Pathfinder test mass (see Sec. \ref{sec:kbub}), which is $2 \mathrm{kg}$ and has a  size of $4.6\mathrm{cm}$~\cite{Armano:2016bkm}. 
From Eq.~\eqref{eq:1P_sol}, the screening radius is $r_\mathrm{sc} \approx 10^{-4} \mathrm{km}$, while $r_\mathrm{UV} \approx 10^{-13} \mathrm{km}$. The $r_\mathrm{UV}$ scale is pushed to even smaller values as $N \gg 1$ (and even further for Galileons~\cite{deRham:2014wfa}).  The presence of a second body will not change this conclusion, see Sec. \ref{sec:k2b2}.

Interestingly, the $r_\mathrm{UV}$ scale for  screening in opposite DBI  is significantly larger:
\begin{eqnarray} \label{eq:r_UV_DBI}
	r^\mathrm{(DBI)}_\mathrm{UV} \approx
10^{5} \mathrm{km}  \,\left( \frac{\alpha}{0.1}\right)^{2/3}   \left( \frac{\Lambda}{\rm{meV}} \right)^{-1}   \left( \frac{m}{M_\odot}\right)^{2/3} \,.
\end{eqnarray}
Thus, as already noticed in Ref.~\cite{Burrage:2014uwa}, the opposite  DBI EFT is not appropriate for describing screening around the Sun $(R_\odot = 7 \times 10^5 \mathrm{km})$, at least for $\Lambda \approx \Lambda_{\mathrm{DE}}$. The case becomes even worse for neutron stars, as $R_\mathrm{NS} \sim 10 \mathrm{km} \ll r^\mathrm{DBI}_\mathrm{UV}$. 

In the case of anti-screening\footnote{We thank the authors of Ref.~\cite{Hertzberg:2022bsb} for pointing out a small error in the previous version of this manuscript.} i.e. $1/2 < N < 1$ [see Eq. \eqref{eq:k_anti}], based on the results of Ref.~\cite{deRham:2014wfa}, the rough condition for the regime of validity is $r_\mathrm{a-sc}
\gg \Lambda^{-1}$, where $r_\mathrm{a-sc}$ is the anti-screening radius. This condition is easily satisfied for any relevant astrophysical scenario. The exception is the standard DBI anti-screening $N= 1/2$, where the classical description breaks down around the anti-screening radius.

%In the case of  For p>1/2 and p<1 it appears that one could be outside of the regime of validity only for very small values of the coupling constant x mass of the object that doesn't apply for astrophysical scenarios. 

%Note that a similar problem occurs for anti-screening, where \eqref{eq:r_UV} applies with $1/2 < N < 1$ in the deep anti-screening regime  Taking $N \approx p=5/6$, which  was the case considered in  many examples in Ref.~\cite{Hertzberg:2022bsb}, implies
%
%\begin{eqnarray} \label{eq:r_UV_asc}
%	r^{p=5/6}_\mathrm{UV} \approx 10^{73} \mathrm{km}  \,\left( \frac{\alpha}{10^{-3}}\right)^{5/2}   \left( \frac{\Lambda}{10^2\rm{eV}} \right)^{-1}   \left( \frac{m}{M_\odot}\right)^{5/2} \,,
%\end{eqnarray}
%
%where we have taken $\alpha$ to be consistent with the Cassini bounds and $\Lambda \sim 10^2\mathrm{eV}$ (which corresponds to the lower bound derived  from several astrophysical observations  in Ref.~\cite{Hertzberg:2022bsb}). It appears  that in none of the examples given in Ref.~\cite{Hertzberg:2022bsb} the non-linear regime is within the regime of validity of the EFT. 
%Indeed, it is hard to imagine one such example/scenario
%for  $p=5/6$, even if one considers much higher strong coupling scales and less massive celestial objects (which would however also reduce the anti-screening radius). The situation is somewhat better for smaller $p \geqsim 1/2$, where  for large $\Lambda$ one can find astrophysical systems in the regime of validity of the EFT.

    %%%%%%%%%%%%%%%%%%%%%%%%%%%%%%%%%%%%%%%%%%%%%%%%%%%%%%%
    \section{Regularized Newtonian/FJBD potential} \label{app:coloumb}
    %%%%%%%%%%%%%%%%%%%%%%%%%%%%%%%%%%%%%%%%%%%%%%

    The leading order energy for a system of  two point particles can be found from  Eq.~\eqref{eq:ener_poly}, when  $N=1$, by integrating by parts to substitute $X$ with the source $T$:
    \begin{equation}
        E_\psi = \frac{\alpha}{2M_{\rm{Pl}}}  [m_a \psi(z_a) + m_b \psi(z_b)] \,.
    \end{equation}
   Substituting $\psi$ [taking $\Lambda \to \infty$ in Eq. \eqref{eq:point_pt_pert_exp}], this expression  diverges due to the self-energies of the two particles. 
   Even classically, however, these 
   self-energy contributions are actually finite
   due to the finite size of the two bodies (for which point particles are just a model valid
   in the IR). If the bodies have a finite size, the total binary energy is therefore regular and reads
    \begin{equation}
        E_{\psi,\varepsilon} =-\frac{1}{4\pi} \left(\frac{\alpha}{M_{\rm{Pl}}} \right)^2  \left[ \frac{m_a^2}{\varepsilon}+ \frac{m_b^2}{\varepsilon} + \frac{m_a m_b}{D} \right]\,,
    \end{equation}
where $\epsilon$ is a regularization parameter of the order of the size of the two bodies.
Note that the self energy contributions $\propto 1/\epsilon$ are constant and thus not observable, as the fifth force is given by the energy's gradient. For instance, in the 
 FJBD case the force reads
    \begin{equation} \label{eq:force_psi}
        \frac{dE_{\psi, \varepsilon}}{da} = \frac{1}{4\pi} \left(\frac{\alpha}{M_{\rm{Pl}}} \right)^2 \frac{m_a m_b}{D^2}\,
    \end{equation}
    which is finite and manifestly independent of the regulator $\varepsilon$.
    
    For  numerical purposes, however, we need to specify concrete a ``UV completion" of the point particle model. The Gaussian source that we use in this work admits an analytical solution for the Poisson equation, i.e.
    \begin{equation} \label{eq:lin_trial}
        \psi_{\sigma_i} = -\frac{ m\alpha  }{4 \pi M_{\mathrm{Pl}}} \mathrm{Erf}\Big(\frac{|\bm{r}-\bm{r}_i|}{\sqrt{2}\sigma_i} \Big) \,.
    \end{equation}
    One of course needs to establish that the results do not depend on the choice of $\sigma$, as the Gaussian distribution above is not a physically motivated ``UV model"\footnote{For example, in order to model a  star,  one would need to solve the Einstein-Klein-Gordon system for a realistic matter equation of state, as done e.g. in Refs.~\cite{Armendariz-Picon:2005oog, terHaar:2020xxb,Bezares:2021yek,Hertzberg:2022bsb}.}. Indeed, we find  that  the relative difference between the  FJBD force for Dirac-delta and  Gaussian sources is less than $1\%$ when $a/\sigma \geqsim 2$.

\subsection{Calculation of the fifth force in the irrotational approximation}  \label{app:5force_calc}

As the integrals in Sec.~\ref{sec:force} have poles for the point-particle source, we have used the aforementioned Gaussian regularization to calculate them. Consider first the integral $I_N(q)$ of the fifth force in the deep screening regime [Eq.~\eqref{eq:5_f_sc}], which depends only on
$q$. As the integrand 
scales as
$\sim \mathsf{r}^{(1+2n)/(1-2n)}$ near infinity,  we can formally identify the screening region with the whole space. Using the Gaussian regularization,  we have $I_N(q) \to \Tilde{I}_N(q,\mathsf{R})$, and we have calculated  $ \Tilde{I}_N(q,\mathsf{R})$ for a few values of $\mathsf{R}$. These results are well described by a functional form $\Tilde{I}_N(q,\mathsf{R})=I_N(q) \mathsf{R}^{p}$
when $\mathsf{R} \ll 1$. Using this fact, 
we can extract $I_N(q)$, verifying that $p \approx 0$ i.e. that our results are independent of the details of the regularization.

 Regarding the calculation of the full energy/force  in the irrotational approximation, the volume integral of Eq.~\eqref{eq:ener_poly_2} can be split as %
\begin{eqnarray}
    \mathcal{E} \approx-  \int d\Omega  \Big[ \int^\mathcal{R}_0  d\mathsf{r} \sum^N_{n=1} \Big(\frac{2n-1}{2n} \Big) \mathcal{X}^n + \frac{1}{2} \int^\infty_\mathcal{R} d\mathsf{r} \mathcal{X}_\Psi \Big] \mathsf{r}^2  \,, \nonumber
\end{eqnarray}
where we have assumed  that $\mathcal{X}=\mathcal{X}(\mathcal{X}_\Psi)$ [given by Eq. \eqref{eq:X_quad_invert} for the quadratic kinetic function], $\mathcal{R}  \gg  c_2 \sqrt{\kappa} \mathsf{D}$ and $d\Omega=\sin{\theta}d\theta d\vartheta$.  The second integral can be found in closed form using Mathematica~\cite{Mathematica}. Differentiating the integrand with respect to $\mathsf{D}$ before fixing the scale $\mathsf{D}=1$, we obtain the  magnitude of the fifth force. As we  perform all calculations with the Gaussian regularization, differentiation with respect to $\mathsf{D}$ and the  integral commute. In order for these results  to be independent of the details of the regularization, one must consider the limit $\mathsf{R} \ll 1$. In practice, we find that already at $D \simeq 4R$ the relative difference between the deep screening approximation and the full calculation is smaller than $1\%$. This difference then gradually increases with $D/R$, because of the worsening of the  deep screening approximation, up to $\sim 50\%$ when $D \approx r_{\mathrm{sc}}$.

    %%%%%%%%%%%%%%%%%%%%%%%%%%%%%%%%%%%%%%%%%%%%%%%%%%%%%%%
    \section{Classical dual  vs. Helmholtz decomposition} \label{app:gaba}
    %%%%%%%%%%%%%%%%%%%%%%%%%%%%%%%%%%%%%%%%%%%%%%

In Ref.~\cite{Gabadadze:2012sm} it was shown that the theory described by Eq.~\eqref{eq:action}, with $\beta=-1$ and $\gamma=0$ and in the decoupling limit of the scalar and tensor degrees of freedom, can be reformulated, at the classical level, as
\begin{eqnarray}\label{eq:act_dual}
	\mathcal{L_{\mathrm{dual}}} = - \frac{1}{2} (\partial \varphi)^2 + \frac{3}{4} \Lambda^{4/3} (\Gamma_\mu \Gamma^\mu)^{2/3} - \Gamma^\mu \partial_\mu \varphi \,,
\end{eqnarray}
where one has introduced the new vector field $\Gamma_\mu$.
%where the new vector field $[\Gamma_\mu]=2$ has been introduced. 
The equations of motion that follow from this action are
\begin{eqnarray}
\Box \varphi +\partial_\mu \Gamma^\mu= - \frac{\alpha}{M_{\rm Pl}} T \,, \label{eq:dual_eom_1}  \\
\Lambda^{4/3} (\Gamma_\mu \Gamma^\mu)^{-1/3} \Gamma_\mu = \partial_\mu \varphi \,.\label{eq:dual_eom_2}
\end{eqnarray}
Using the latter, the auxiliary vector $\Gamma_\mu$ can be integrated out and the original action is recovered. The appeal of this formulation is that none of the coupling constants in the dual action given by Eq.~\eqref{eq:act_dual} have a negative mass dimension (except for the scalar-matter coupling). This in turn allows for a controlled perturbative expansion in the non-linear regime of the original theory. This formulation was shown to originate from a Legendre transformation and is generalizable to a large class of self-interacting theories~\cite{Padilla:2012ry}.

Let us now Hodge-Helmholtz decompose the vector and redefine the scalar as
\begin{eqnarray} \label{eq:dual_dec}
	\Gamma_\mu=(-2)B_\mu + \partial_\mu \Tilde{\Gamma} \,, \quad \partial_\mu B^\mu=0   \,, \quad \varphi = \psi-\Tilde{\Gamma} \,,
\end{eqnarray}
where $B_\mu,\psi$ are for now generic objects. Substituting this decomposition into Eq. \eqref{eq:dual_eom_1} we obtain Eq. \eqref{boxpsi}, i.e.
\begin{eqnarray}
   \Box \psi = -\frac{1}{2M_{\rm Pl}} T \,.
\end{eqnarray}
Squaring Eq.~\eqref{eq:dual_eom_2} and substituting the above decomposition, we reconstruct Eq. \eqref{eq:helmh_4v} for the quadratic $k$-essence, i.e.
\begin{eqnarray}
- \frac{1}{2} \Big(1 + \frac{X}{\Lambda^4} \Big) \partial_\mu \varphi = -\frac{1}{2}\partial_\mu \psi + B_\mu \,.
\end{eqnarray}
Thus, the dual formulation of quadratic $k$-essence is equivalent to the Hodge-Helmholtz decomposition.

In Ref.~\cite{Gabadadze:2012sm} a different decomposition was used instead of Eq.~\eqref{eq:dual_dec}:
\begin{eqnarray}
	&& \Gamma_i=(-2)B_i + \partial_i \Tilde{\Gamma} \,, \quad  \Gamma_0=\omega  \,, \quad \partial_i B^i =0 \,, \quad \\
	&& \varphi = \psi-\Tilde{\Gamma}  \,. \nonumber
\end{eqnarray}
It is easy to show that in the static regime,  this decomposition is also equivalent to the $\mathbb{R}^3$ Helmholtz one in  Eq.~\eqref{eq:helmh_3v}. It was then argued in Ref.~\cite{Gabadadze:2012sm} that one can consistently choose an ansatz where $B_i=0$. This however is not the case, as we have elaborated in the main body of this work. The reason why this inconsistency has not been noticed in Ref.~\cite{Gabadadze:2012sm} is that the dual formulation was applied to isolated systems in spherical and cylindrical symmetry, where the solenoidal component vanishes as argued in Sec. \ref{sec:EoM_R3}, \ref{sec:an_sph}.

Finally, let us note an advantage of the Helmholtz-decomposition program over the perturbative expansion in $\Tilde{\Gamma}$ that was performed in Ref.~\cite{Gabadadze:2012sm}, for scenarios where the solenoidal component is zero. Instead of expanding,  one can solve for the square of the scalar gradient  to all orders in the perturbative expansion (see Sec. \ref{sec:analy}). If one is interested in the fifth force or the force acting on  test bodies, the scalar gradient is the relevant object and the full scalar profile is not necessary (Sec. \ref{sec:force}). However, the dual formulation at the level of action can have other advantages, e.g. allowing for constructing analogues of the irrotational approximation in other types of theories, like Galileons.
 
    %%%%%%%%%%%%%%%%%%%%%%%%%%%%%%%%%%%%%%%%%%%%%%%%%%%%%%%
    \section{Code validation} \label{app:num_test}
    %%%%%%%%%%%%%%%%%%%%%%%%%%%%%%%%%%%%%%%%%%%%%%

    \begin{table*}
   	 \begin{tabular}{||c c c c c c c||}
   		 \hline
   		 & $\kappa$ & $q$  & $\mathsf{R}$ & $\mathsf{D}/\mathsf{R}$  & $\varrho_\mathrm{fin}$  & $\mathsf{z}_\mathrm{fin}$ 	\\[0.5ex]
   		 \hline\hline
   		 \hline
   		 (i) & 6 & $\infty$ & 1.6 & 4  & 6 & 12 \\
   		 \hline
   		 (ii) & 4.8 & 1 & 1.6 & 1.25 & 8.4 & 23.04   \\
   		 \hline \end{tabular} \caption{Parameters of the systems considered for the numerical tests, in units given by Eq.~\eqref{eq:tildas_units}.} \label{tab:app_num}
    \end{table*}
    
    We start by testing our code
against the known solution  for a single isolated body. Because our Newton-Raphson/Broyden 
method needs an initial guess for the solution, we start from the ``linear'' solution given by Eq.~\eqref{eq:lin_trial}, while the exact solution to which we compare is known analytically up to 
integration of the  ordinary differential equation gievn by Eq. \eqref{eq:X_quad_invert}.
As  a test of our non-linear elliptic solver, we have then solved numerically for the one-body system (i)  given in Tab. \ref{tab:app_num}, for different grid resolutions. 

\begin{figure}[h]
   	 \centering
   	 \includegraphics[width=0.48\textwidth]{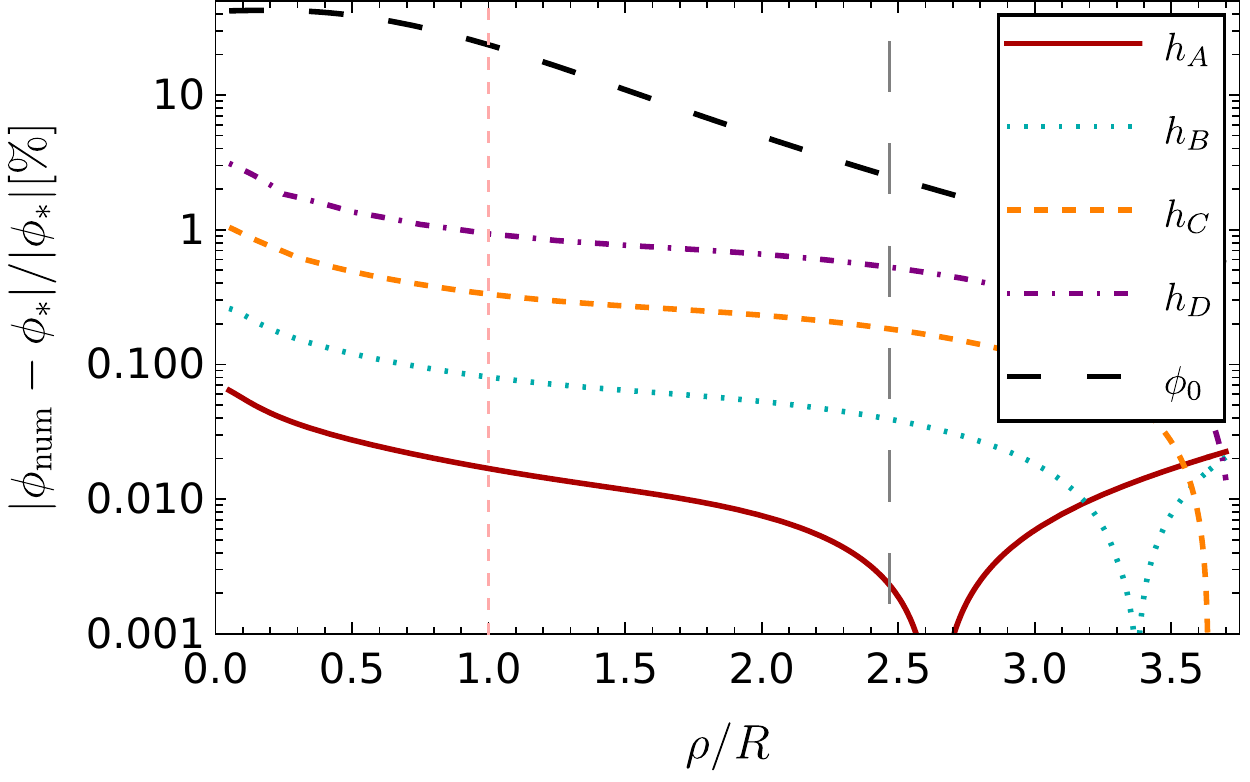}
   	 \caption{
     Relative difference of our numerical results $\phi_{\rm num}$ from the semi-analytic solution $\phi_\ast$, for a one-body system and
 four different resolutions $h/R=\{0.0375,0.075,0.15,0.25\}$ (corresponding respectively to $A, B, C, D$), at the center of the source and as function of $\varrho$. The black long-dashed line represents the difference between the initial guess (FJBD scalar sourced by a Gaussian) and the semi-analytic solution.  The screening radius $r_{\rm sc}=2.47 R $ is shown by a gray, long dashed line, and the effective radius of the Gaussian source  is $R=2\sigma$ (pink, short dashed line). }
   	 \label{fig:num_rel_1p}
    \end{figure}

In Fig. \ref{fig:num_rel_1p}, we show the relative difference between the numerical and the semi-analytic results as a function of $\varrho$, 
with $\mathsf{z}=\mathsf{z}_c$ fixed to the center of the matter source (where the numerical error is the largest). Results for grid resolutions $\mathsf{h}/R \leqsim 0.15$ have sub-percent errors with respect to the semi-analytic ones. The plot also shows that the relative difference between the initial guess, i.e. the FJBD scalar sourced by a Gaussian [Eq.~\eqref{eq:lin_trial}], and the semi-analytic solution that be as large as $\sim30\%$. Finally, we have checked that the solenoidal vector $\bm{B}$ is zero (Sec. \ref{sec:an_sph}), up to a numerical error.

\begin{figure*}[th]
\begin{tabular}{cc}
\includegraphics[width=.48\textwidth]{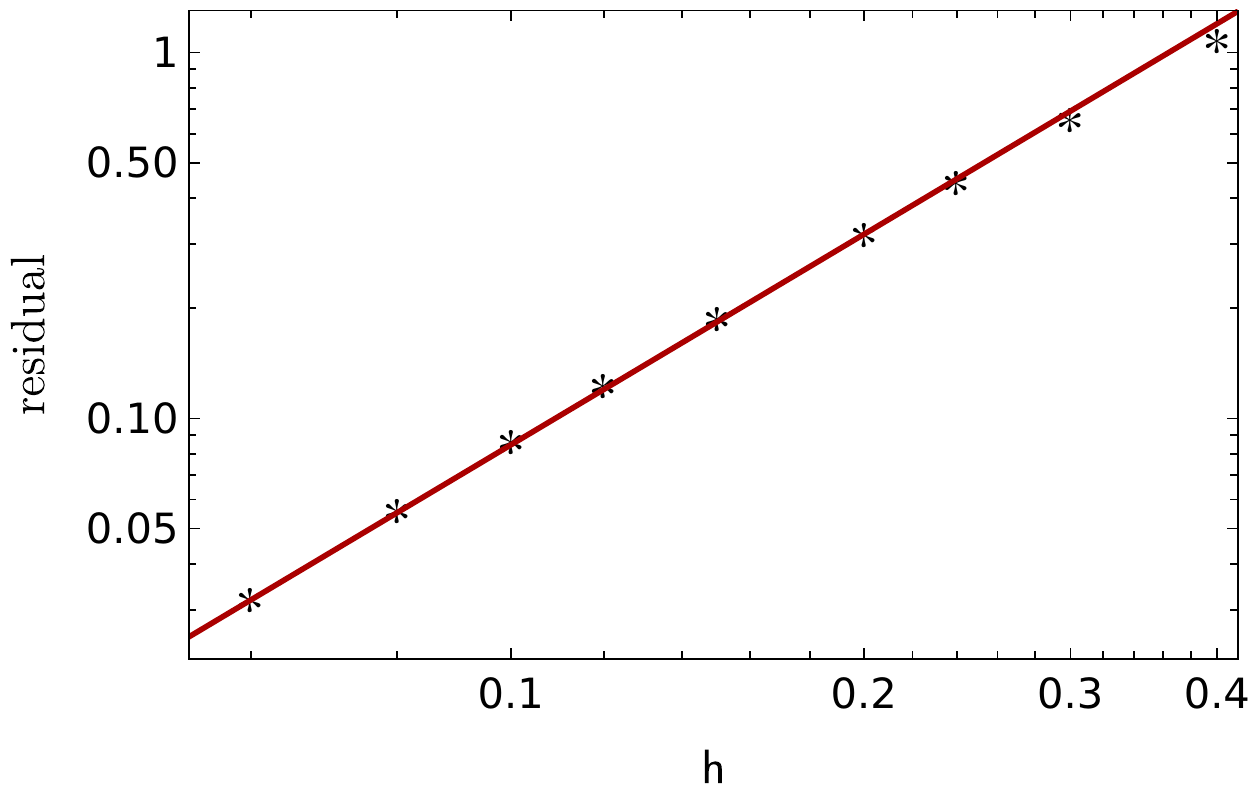}
\qquad
\includegraphics[width=.48\textwidth]{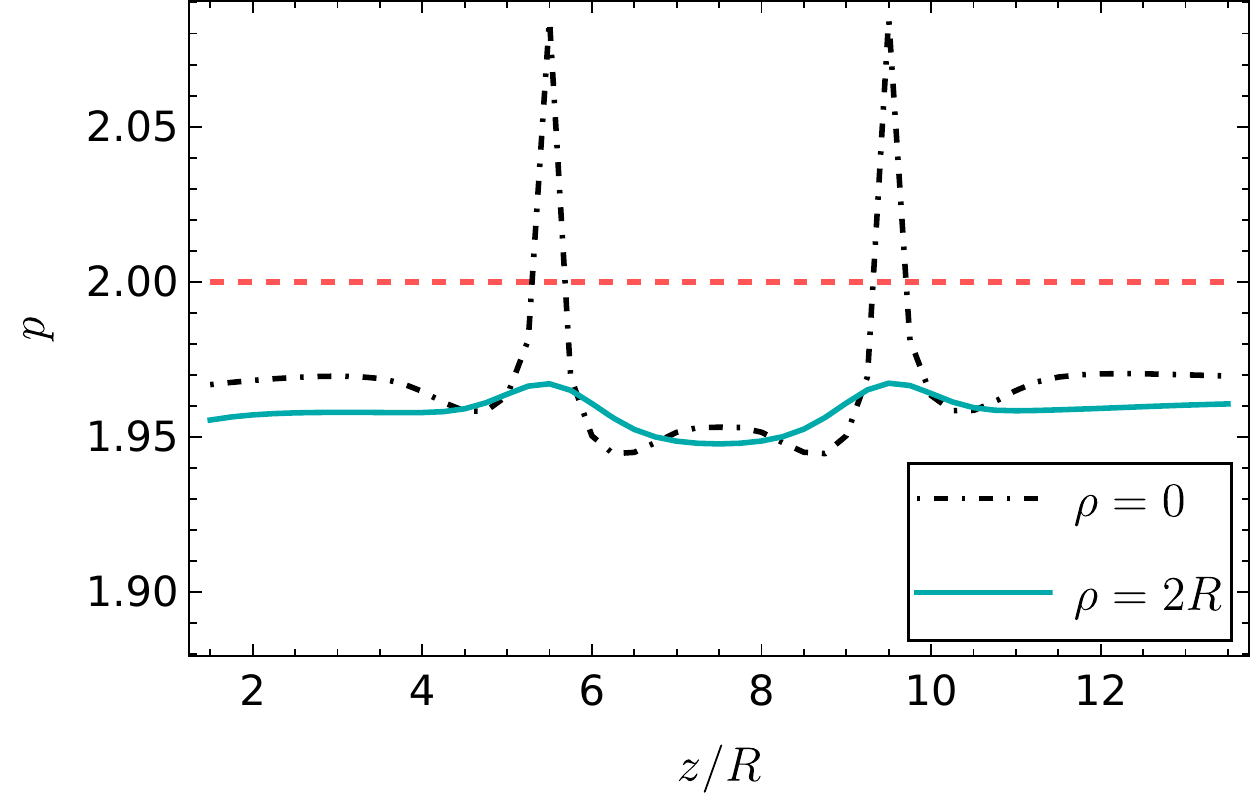}
\end{tabular}
\caption{Grid resolution convergence tests: (left) $L^2$ norm of the residuals from the semi-analytic solution for a single isolated object (i) in Table \ref{tab:app_num}, as a function of the resolution. The fitted power law (solid line) corresponds to an exponent  $p=1.91$; (right) Effective convergence order $p$ for the two-body system (ii) in Table \ref{tab:app_num}, evaluated at $\rho=0$ and $\rho=2R$ and as function of $z$, is consistent with the implemented second order convergence scheme.}
	\label{fig:2p_conv}
\end{figure*}

We have also checked the convergence of our residuals (from the semi-analytic solution). In more detail, 
in Fig. \ref{fig:2p_conv} (left) we show the $L^2$ norm
of the residuals (throughout the grid) vs the grid resolution, alonside a power law fit (red line). The fitted power law exponent ($p=1.91$) is very close to $p=2$, as expected from our discretization scheme. By using instead the $L^1$ norm of the residuals, we obtain $p=1.99$.

    In the two-body case, we do not have a semi-analytic exact solution to compare our numerical results with. However, we can test convergence
    by considering three different resolutions ($\mathsf{h}_1=0.32, \mathsf{h}_2=0.16, \mathsf{h}_3=0.08$), and estimating the convergence order as 
\begin{eqnarray}
    p=\log_2\left(\frac{|\phi_1-\phi_3|}{|\phi_2-\phi_3|} -1\right)
\end{eqnarray}
    where ${\phi}_1$, ${\phi}_2$ and ${\phi}_3$
    are the numerical solutions.
    	For the scenario (ii) in Tab. \ref{tab:app_num},  Fig. \ref{fig:2p_conv} (right)
     shows $p(\varrho,\mathsf{z})$ at two radial points and demonstrates that the results are consistent with the expected second order convergence.

    \bibliography{references}
\end{document}